\documentclass[aps,twocolumn,prd,showpacs,nofootinbib]{revtex4}

\usepackage{amsmath}
\usepackage{graphicx}
\usepackage{subfigure}
\usepackage{dcolumn}
\usepackage{bm}
\usepackage{amssymb}
\usepackage{latexsym}
\usepackage{epsf}
\usepackage{graphics}
\usepackage{amssymb}
\usepackage{amsfonts}
\usepackage[mathscr]{eucal}
\usepackage{amsmath}
\usepackage{enumerate}
 \usepackage{hhline}
\usepackage{array}
\usepackage{tabularx}

\newcommand{\lta}{\lesssim}
\newcommand{\gta}{\gtrsim}

\newcommand{\dd}{\mathrm{d}}

\newcommand{\nS}{n_{_{\mathrm{S}}}}

\newcommand{\mean}[1]{\left\langle #1 \right\rangle}
\newcommand{\mpl}{m_{_\mathrm{Pl}}}

\newcommand{\ie}{{\it i.e.}~}
\newcommand{\eg}{{\it e.g.}~}

\newcommand{\bear}{\begin{array}}  \newcommand{\eear}{\end{array}}
\newcommand{\bef}{\begin{figure}}  \newcommand{\eef}{\end{figure}}
\newcommand{\bec}{\begin{center}}  \newcommand{\eec}{\end{center}}
  
\newcommand{\lmk}{\left(}  \newcommand{\rmk}{\right)}
\newcommand{\lkk}{\left[}  \newcommand{\rkk}{\right]}

\newcommand{\cl}{\mathrm{cl}}
\newcommand{\ini}{\mathrm{in}}

\newcommand{\df}{\delta\varphi}

\newcommand{\e}{\mathrm{e}}

\newcommand{\IR}{_{\mathrm{IR}}}
\newcommand{\UV}{_{\mathrm{UV}}}

\begin{document}

\title{Geometrically Consistent Approach to Stochastic DBI Inflation}

\author{Larissa Lorenz} \email{larissa.lorenz@uclouvain.be}
\affiliation{Theoretical and Mathematical Physics Group, Centre for Particle Physics and Phenomenology,
Louvain University, 2 Chemin du Cyclotron, 1348 Louvain-la-Neuve, Belgium}
\author{J\'er\^ome Martin} \email{jmartin@iap.fr} \affiliation{
Institut d'Astrophysique de Paris, UMR 7095-CNRS, Universit\'e Pierre
et Marie Curie, 98bis boulevard Arago, 75014 Paris, France}

\author{Jun'ichi Yokoyama} \email{yokoyama@resceu.s.u-tokyo.ac.jp}
\affiliation{Research Center for the Early Universe, Graduate School
  of Science, The University of Tokyo, Tokyo 113-0033, Japan}
\affiliation{Institute for the Physics and Mathematics of the Universe
  (IPMU), The University of Tokyo, Kashiwa, Chiba, 277-8568, Japan }

\date{\today}

\begin{abstract}
  Stochastic effects during inflation can be addressed by averaging
  the quantum inflaton field over Hubble-patch sized domains. The
  averaged field then obeys a Langevin-type equation into which
  short-scale fluctuations enter as a noise term. We solve the
  Langevin equation for a inflaton field with Dirac Born Infeld (DBI)
  kinetic term perturbatively in the noise and use the result to
  determine the field value's Probability Density Function (PDF).  In
  this calculation, both the shape of the potential and the warp
  factor are arbitrary functions, and the PDF is obtained with and
  without volume effects due to the finite size of the averaging
  domain. DBI kinetic terms typically arise in string-inspired
  inflationary scenarios in which the scalar field is associated with
  some distance within the (compact) extra dimensions. The inflaton's
  accessible range of field values therefore is limited because of the
  extra dimensions' finite size. We argue that in a consistent
  stochastic approach the distance-inflaton's PDF must vanish for
  geometrically forbidden field values. We propose to implement these
  extra-dimensional spatial restrictions into the PDF by installing
  absorbing (or reflecting) walls at the respective boundaries in
  field space. As a toy model, we consider a DBI inflaton between two
  absorbing walls and use the method of images to determine its most
  general PDF.  The resulting PDF is studied in detail for the example
  of a quartic warp factor and a chaotic inflaton 
  potential. The presence of the walls is shown to affect the inflaton
  trajectory for a given set of parameters.
\end{abstract}

\maketitle

\section{Introduction}
\label{sec:intro}

Putting the successful inflationary scenario on the firm footing of a
fundamental theory is one of the remaining challenges in
cosmology. Recent years have seen considerable progress towards this
goal with the construction of several concrete string inflation
models; for recent reviews, see
\eg~\cite{McAllister:2007bg,Kinney:2009vz,Baumann:2009ni,Lorenz:2010nx}
and references therein. A top-level distinction among these models is
the either closed or open string mode character of the inflaton
field. A typical example of the second class are brane inflation
scenarios~\cite{Dvali:1998pa,Burgess:2001fx,Kachru:2003sx}: the
inflaton field $\phi$ corresponds (up to renormalization) to the
distance between two branes embedded in a higher-dimensional
background. While there is an ongoing debate about the form of the
inflationary potential
$V(\phi)$~\cite{Kachru:2003sx,Baumann:2006th,Baumann:2007np,Baumann:2007ah,
  Krause:2007jk,Pajer:2008uy,Baumann:2008kq}, a generic feature of
these models is the field's kinetic term, which is of Dirac Born
Infeld (DBI) type rather than
canonic~\cite{Silverstein:2003hf,Alishahiha:2004eh}. One can
understand these DBI dynamics as a geometry-imposed upper limit on the
field's velocity $\dot{\phi}$, characterized by the so-called warp
factor $T(\phi)$. This relativistic speed limit acts like a brake on
the inflaton, forcing $\phi$ to ``slow-roll'' even in regions where
the potential is not flat. Hence, open string mode inflaton models
provide an additional mechanism to generate quasi-exponential
expansion.

\par

It is interesting to investigate the DBI analogues of standard
inflationary calculations, such as the field perturbations' evolution
and spectra~\cite{Lorenz:2008je,Lorenz:2008et}, but also the effects
of stochastic
inflation~\cite{Starobinsky:1986fx,Starobinsky:1994bd,Martin:2005ir,
  Martin:2005hb,Kuhnel:2008yk,Finelli:2008zg,Kuhnel:2008wr,Seery:2009hs,
  Kuhnel:2010pp,Finelli:2010sh}. Stochastic inflation provides a
technique to assess quantum effects on the inflaton $\phi$'s
trajectory averaged over a scale beyond the Hubble patch. In this way,
one can define a coarse-grained field $\varphi$ which consists
exclusively of the large-scale Fourier components. To zeroth order,
$\varphi$ obeys the slow-roll Klein Gordon equation, but its full
evolution is subject to stochastic noise $\xi$ from small-scale
Fourier modes. For standard inflation with arbitrary potentials, this
equation was solved perturbatively up to $\mathcal{O}(\xi^{2})$ in
Ref.~\cite{Martin:2005ir}. Using this perturbative solution,
Ref.~\cite{Martin:2005ir} showed how to obtain the Probability Density
Function (PDF) for $\mean{\varphi}$ in the Gaussian approximation,
with and without the volume effects (\ie the size of each averaging
domain) taken into account. The reliability of this treatment was
further studied in Ref.~\cite{Martin:2005hb}. In this paper, we
generalize both of these results to the case of DBI inflation with
arbitrary warp factors and potentials.

\par

The Langevin equation for DBI models was given for the first time in
Ref.~\cite{Chen:2006hs}, which aimed at studying ``eternal
inflation''~\cite{Vilenkin:1983xq,Linde:1986fd} in the brane inflation
context. Since then, stochastic effects in DBI (or general
$k$-inflation) models have been studied by several
authors~\cite{Helmer:2006tz,Tolley:2008na}. In this paper, we use the
approach of Refs.~\cite{Martin:2005ir, Martin:2005hb} to calculate the
PDFs of a DBI inflaton field (with and without volume effects) for
arbitrary functional form of the potential and warp factor. We
illustrate our results by applying them to the example model of
``chaotic Klebanov Strassler (CKS) inflation'', where $T(\phi)$ and
$V(\phi)$ are known functions. The integrals encountered are, within
certain limits, exactly calculable, and we discuss the behavior and
reliability of the resulting DBI inflaton probability densities.

\par

However, we argue that all PDFs used so far in the literature suffer
from a serious problem: they predict a non-vanishing probability for
the moving $D3$-brane to find itself outside the so-called ``Klebanov
Strassler (KS) throat'', \ie outside the (part of the)
extra-dimensional geometry with warp factor $T(\phi)$. In fact, this
problem is twofold. Firstly, at the bottom end of the throat 
this means that there is a
non-vanishing probability to literally find 
the brane ``outside'' the
extra dimensions, in other words ``out of space'', which is clearly
meaningless. Secondly, since the metric of
the 6d bulk space [and hence the continuation of $T(\phi)$] is
typically unknown beyond the KS region, a string-inflationary scenario
based on the brane's motion inside the throat becomes inconsistent
beyond its top end. Note that, while the latter question is 
of a more technical nature and we may hope to resolve it as our 
understanding of string geometries improves, the former issue 
is rather severe as the inflaton's PDF does not respect 
the fundamentally geometric origin of the scenario.

\par
Put a
different way, the compact character of the stringy extra dimensions
(for the purposes of concrete model building, this means the
well-known KS ``corner'' of the 6d manifold) can be translated
directly into a restricted field range for the DBI inflaton. For
consistency, any modifications of the classical trajectory induced by
stochastic effects should still respect these geometry-imposed
boundaries in field space. Hence, studies of stochastic DBI inflation
so far were missing a tool to ensure the consequences 
of a stringy inflaton's geometric interpretation 
at the effective field theory level.

\par

To amend this problem, we propose to install ``walls'' at the
boundaries (\ie the bottom $r_{0}$ and the edge $r_{\UV}$) of the
inflationary KS throat (whose radial coordinate is denoted by $r$).
As a consequence, the stochastically corrected inflaton field value
should remain within its allowed range $\phi_{0}<\phi<\phi_{\UV}$ at
all times. This requires the calculation of a new PDF respecting the
boundary conditions imposed by the presence of the reflecting or
absorbing walls. For the first time, we then determine this PDF in the
presence of two absorbing walls at $\phi_{0}$ and $\phi_{\UV}$, using
the method of images, see \eg 
\cite{Sommerfeld:ED, Sommerfeld:PDEs}.  As a by-product, we obtain
the modified stochastic trajectory of the mobile brane within the KS
throat and show that, in some cases, the presence of the walls has a
significant effect.

\par

This paper is organized as follows. In the next Section, we start from
the underlying background equations of DBI inflation and discuss how
they can be used to formulate their stochastic counterpart, \ie the
DBI Langevin equation. We pay special attention to the normalization
factor of the noise term. We then solve the DBI Langevin equation up
to second order in the noise and calculate the corresponding PDF along
with its volume correction. In Sec.~\ref{sec:applications}, we use
these expressions to calculate the stochastic effects in the case of
chaotic Klebanov Strassler inflation, both for a potential with and
without a constant term. Where applicable, we calculate the domain of
validity of the perturbative approach and consider the existence of a
regime of eternal inflation~\cite{Vilenkin:1983xq,Linde:1986fd}.  In
Sec.~\ref{sec:finitesize}, we use the additional information on the
inflaton's geometric bounds as an argument to implement two absorbing
walls into the calculation of the PDF. In this way, it can be assured
that even quantum effects do not violate these field space limits.
Finally, we summarize our main findings in Sec.~\ref{sec:conclusions},
commenting on eternal inflation as well as on an overall picture for a
realistic brane trajectory across the entire compact 6d geometry.

\section{DBI Langevin Equation}
\label{sec:DBI}

\subsection{DBI Background Equations}
\label{subsec:dbiclassical}

As a first step, we quickly recall the basic equations of DBI
inflation at the classical level. Since models of this kind descend
from (\eg type IIB) string theory, they are originally represented by
a higher-dimensional ($d=10$) action for the stringy background and
the embedded branes. After compactification to four dimensions, the
effective field theory typically contains a gravity sector described
by General Relativity and a four-dimensional inflaton field
$\phi\left({\bm x},t\right)$ corresponding to the inter-brane distance
along one of the compactified dimensions. The model's effective 4d
action therefore reads
\begin{eqnarray}
\label{eq:actionDBI}
S &=& -\int {\rm d}^4x\sqrt{-g}
\Biggl[\frac{R}{2\kappa}+V(\phi)-T(\phi)
\nonumber \\ & & +T(\phi)
\sqrt{1+\frac{1}{T(\phi)}g^{\mu \nu}\partial _{\mu}\,\phi 
\partial _{\nu}\phi}
\Biggr]\, ,
\end{eqnarray}
where $R$ is the four-dimensional scalar curvature and $\kappa
=8\pi/\mpl^2$, $\mpl$ being the Planck mass. It is useful to introduce
(making use of the analogy with Special Relativity) the so-called
``Lorentz factor'' $\gamma(\phi,\partial_{\mu}\phi)$, defined as
\cite{Alishahiha:2004eh}
\begin{equation}\label{eq:defgammageneral}
\gamma(\phi,\partial_{\mu}\phi)=
\frac{1}{\sqrt{1+g^{\mu \nu}\partial _{\mu}\,\phi 
\partial _{\nu}\phi/T(\phi)}}\,.
\end{equation}
Roughly speaking, this Lorentz factor measures how close the inflaton
field's velocity is to the geometry-imposed speed limit
$\sqrt{T(\phi)}$ (see below).  In terms of $\gamma$ (the arguments of
which we frequently suppress below), one can rewrite
Eq.~(\ref{eq:actionDBI}) in the simple form
\begin{equation}
S = -\int {\rm d}^4x\sqrt{-g}
\left[\frac{R}{2\kappa}+V(\phi)-
\frac{\gamma-1}{\gamma}\,T(\phi)\right]\,.
\end{equation}

\par

Specifying to a Friedmann Lema\^{\i}tre Robertson Walker (FLRW)
universe with homogeneous scalar field matter, the Friedmann and Klein
Gordon equations read
\begin{eqnarray}
H^{2}&=&\frac{\kappa}{3}\left[(\gamma-1)T(\phi)+V(\phi)\right]\, ,
\label{eq:Hdbi}\\
-\frac{V'(\phi)}{\gamma^{3}}&=&\ddot{\phi}+\frac{3H}{\gamma^{2}}
\dot{\phi}+\frac{3\gamma-\gamma^{3}-2}{2\gamma^{3}}\,T'(\phi)\, 
,\label{eq:eofmdbi}
\end{eqnarray}
where a prime denotes a derivative with respect to the inflaton field
$\phi$. Note that in the FLRW case, $\gamma$ from
Eq.~(\ref{eq:defgammageneral}) simplifies to
\begin{equation}
\label{eq:gamma}
\gamma(\phi,\dot{\phi})=\frac{1}{\sqrt{1-{\dot{\phi}}^{2}/T(\phi)}}\, ,
\end{equation} 
where the notion of $\sqrt{T(\phi)}$ as a speed limit is evident.
Using the square root's expansion in Eq.~(\ref{eq:gamma}), it is clear
from Eq.~(\ref{eq:eofmdbi}) that, while $\gamma\approx1$, the inflaton
is close to standard dynamics. In this regime, $\phi$ obeys the usual
Klein Gordon equation and hence can slow-roll only if the potential is
sufficiently flat. Using the full expression (\ref{eq:gamma}),
however, one can see that the inflaton velocity can never exceed
$\sqrt{T(\phi)}$ even if the potential is steep. The limit where
$\dot{\phi}\rightarrow\sqrt{T(\phi)}$ (and hence
$\gamma\rightarrow\infty$) can therefore be thought of as an
additional, ``ultra-relativistic'' regime of inflation. Let us make
this statement more precise by taking the time derivative of the DBI
Friedmann equation~(\ref{eq:Hdbi}): the DBI condition to maintain
accelerated expansion reads
\begin{equation}
\label{eq:DBIcondition}
\frac{\ddot{a}}{a}=\frac{\kappa}{3}\,V(\phi)
-\frac{\kappa}{6}\frac{(\gamma-1)(\gamma+3)}{\gamma}\,T(\phi)>0\,.
\end{equation}
From this expression it is evident that the potential $V(\phi)$ still
has to dominate the energy density.

\par

Our next step is to notice that from combining Eqs.~(\ref{eq:Hdbi})
and~(\ref{eq:eofmdbi}) one obtains
\begin{equation}
\label{eq:phidotdbi}
\dot{\phi}=-\frac{2H'}{\kappa\gamma} \, .  
\end{equation}
This formula has two important consequences. Firstly, replacing
$\dot{\phi}$ by this expression in Eq.~(\ref{eq:gamma}), the Lorentz
factor $\gamma$ is easily expressed as a function of $\phi$ only
\cite{Chen:2006hs},
\begin{equation}
\label{eq:gammaphi}
\gamma(\phi)=\sqrt{1+\frac{4H'^{2}}{\kappa^{2}T(\phi)}}\, . 
\end{equation}
Secondly, using the DBI slow-roll condition $H^{2}\approx \kappa
V(\phi)/3$, which is derived and justified in detail in
Appendix~\ref{app:srlimit}, one obtains the following first order
differential equation for $\phi$:
\begin{equation}\label{eq:KG-DBIsr}  
\dot{\phi}\approx-\frac{V'(\phi)}{3\gamma (\phi)H(\phi)}
\end{equation}
Obviously, except for the factor $\gamma $ appearing in the the
denominator, this is the standard Klein Gordon equation in the
slow-roll limit. A detailed derivation of Eq.~(\ref{eq:KG-DBIsr}) is
given in Appendix~\ref{app:srlimit}.
 
\subsection{Stochastic DBI Inflation}
\label{subsec:stochainf}

We now proceed to applying the stochastic approach of
Ref.~\cite{Martin:2005ir} to DBI inflation. In a first step, the
``classical'' inflaton field $\phi(t)$ is replaced by the
coarse-grained field $\varphi(t)$, which is a stochastic
process. Based on the previous considerations [see
Eq.~(\ref{eq:phidotdbi})], we expect $\varphi $ to obey a Langevin
equation of the form
\begin{equation}
\dot{\varphi}=-\frac{2}{\kappa}\frac{H'}{\gamma}
+\mathcal{C}\xi(t)\, ,
\label{eq:DBInoiseadded}
\end{equation}
where $\xi(t)$ is a noise term, describing the short wavelength part
of the full inflaton field, and obeying the following properties,
\begin{equation}
\mean{\xi(t)}=0\,,\qquad\mean{\xi(t)\xi(t')}=\delta(t-t')\,.
\label{eq:noise}
\end{equation}
To proceed from here, the crucial question is how to determine the
normalization factor ${\cal C}$.

\par

In standard (non-DBI) inflation, this question may be resolved in
several ways. One possible route to follow is to normalize the
prefactor $\mathcal{C}$ according to the two-point correlation
function of a massless test field in a de Sitter background. Indeed,
in this case, the above Langevin equation (\ref{eq:DBInoiseadded})
(setting $\gamma =1$) reduces to $\dot{\varphi}={\cal C}\xi(t)$, with
its trivial solution $\varphi(t)={\cal C}\int_{t_\ini}^t\xi(\tau){\rm
  d}\tau$. Using Eqs.~(\ref{eq:noise}), one finds
\begin{equation}
\mean{\varphi ^2(t)}={\cal C}^2\left(t-t_\ini\right)\, .
\end{equation}
Since the exact result is known and reads $\mean{\varphi ^2(t)}=H^3
t/(4\pi^2)$ (setting $t_\ini=0$), we can read off immediately that
${\cal C}=H^{3/2}/(2\pi)$. Another way to see this is to interpret
Eq.~(\ref{eq:DBInoiseadded}) as a Brownian
motion~\cite{Vilenkin:1983xq}, the field undergoing quantum kicks of
amplitude 
%$\sqrt{\mean{\phi_k^2}}=H/(2\pi)$ in every Hubble time
$H/(2\pi)$ in every Hubble time
interval $H^{-1}$.  This leads to the expression
$\mean{\varphi^2(t)}=[H/(2\pi)]^2\,n$, with the number of steps $n$
given by $n=t/H^{-1}=Ht$.

\par

In the case at hand here, \ie stochastic inflation with a DBI inflaton
field, we need to reproduce the same considerations, but for the
modified Langevin equation (\ref{eq:DBInoiseadded}) that now comprises
the Lorentz factor $\gamma$. For this purpose, let us remark that the
fully covariant formulation of the DBI Klein Gordon equation derived
from the action Eq.~(\ref{eq:actionDBI}) reads
\begin{widetext}
\begin{eqnarray}
\label{eq:eomcovariant}
g^{\mu\nu}\nabla_{\mu}\nabla_{\nu}\phi-
\frac{\gamma^{2}}{T}\,g^{\alpha\beta}g^{\mu\nu}
\left(\nabla_{\alpha}\nabla_{\mu}\phi\right)\,
\nabla_{\beta}\phi\,\nabla_{\nu}\phi-
\frac{V'}{\gamma}+\frac{T'}{2\gamma}\,\left(\gamma^{3}-3\gamma+2\right)&=&0.
\end{eqnarray}
To see that this gives back Eq.~(\ref{eq:eofmdbi}) for spatially
homogeneous DBI field $\phi=\phi(t)$ in a FLRW universe, use the
definition (\ref{eq:defgammageneral}) [or (\ref{eq:gamma}),
respectively] of the Lorentz factor $\gamma$ and note that
\begin{equation}
g^{\alpha\beta}g^{\mu\nu}\left(\nabla_{\alpha}\nabla_{\mu}\phi\right)\,
\nabla_{\beta}\phi\,\nabla_{\nu}\phi=\frac{\gamma^{2}-1}
{\gamma^{2}}\,T\ddot{\phi}.
\end{equation}
At the perturbed level,  setting 
\begin{equation}
  \phi({\bm x},t)=\phi(t)+\delta\phi({\bm x},t)\,,\qquad\dd s^{2}
  =-\left(1+2\Phi\right)\dd t^{2}+a^{2}(t)\left(1-2\Phi\right)
  \delta_{ij}\dd x^{i}\dd x^{j}\,,
\end{equation}
one can show through repeated use of the Einstein equations for a DBI
scalar field (see Appendix \ref{app:dphiderivation}) that
\begin{eqnarray}
\label{eq:dphifinal_dbi}
&&\delta\ddot{\phi}_{\bm k}+3H\left(1+\delta_{1}\right)\delta\dot{\phi}_{\bm k}
  +H^{2}\biggl(\frac{k^{2}}{a^{2}H^{2}\gamma^{2}}+2\epsilon_{1}
  -\frac{3}{2}\,\epsilon_{2}-2\epsilon_{1}^{2}-
  \frac{\epsilon_{2}^{2}}{4}
    +\frac{5}{2}\,\epsilon_{1}\epsilon_{2}
  -\frac{\epsilon_{2}\epsilon_{3}}{2}+\frac{3}{2}\,
  \delta_{1} +\frac{3}{2}\,\delta_{1}\epsilon_{1}-\delta_{1}
  \epsilon_{2}
\nonumber \\ & & 
+\frac{5}{4}\,\delta_{1}^{2}+
  \frac{\delta_{1}\delta_{2}}{2}\biggr)\delta\phi_{\bm k} 
=
\frac{H^{2}z}{a\gamma^{3/2}}\,\Phi_{\bm k}\left(2-2\epsilon_{1}+
    \epsilon_{2}+2\delta_{1}\right)
\end{eqnarray}
\end{widetext}
in Fourier space with comoving wavenumber ${\bm k}$. Here, we have
used the $(\epsilon_{i},\delta_{i})$ parameters defined in Appendix
\ref{app:srlimit} to write Eq.~(\ref{eq:dphifinal_dbi}) in a compact
form. Note that in the limit where $\gamma\rightarrow 1$,
$\delta_{i}\rightarrow0$, this gives back precisely the same equation
as in the standard case (see Appendix \ref{app:dphiderivation}).

\par

Ignoring the metric perturbation $\Phi_{\bm k}$ amounts to dropping
the last term in Eq.~(\ref{eq:dphifinal_dbi}).  Moreover, in the ``DBI
slow-roll'' regime, all the $(\epsilon_{i},\delta_{i})$ parameters are
small.  Hence, if in addition we neglect all terms of at least linear
order in these parameters in Eq.~(\ref{eq:dphifinal_dbi}), we obtain
the limit
\begin{equation}
\label{eq:pertDBIapprox}
\delta\ddot{\phi}_{\bm k}+3H\delta\dot{\phi}_{\bm k}
+\frac{k^{2}}{a^{2}\gamma^{2}}\,\delta\phi_{\bm k}\approx 0\, ,
\end{equation}
which again precisely corresponds to the standard equation up to the
replacement $k\rightarrow k/\gamma$.

\par

Usually, however, the perturbed Klein Gordon equation is not written
as in Eq.~(\ref{eq:dphifinal_dbi}), but in terms of the
Mukhanov-Sasaki variable $v_{\bm k}$, which in the DBI case is defined
as the combination
\begin{equation}
\label{eq:trafo}
v_{\bm k}= a\gamma^{3/2}\,\delta\phi_{\bm k}+z\Phi_{\bm k},
\end{equation}
where $z=a\gamma\sqrt{\epsilon_{1}}$. It can be shown (see Appendix
\ref{app:dphiderivation}) by inserting this definition into
Eq.~(\ref{eq:dphifinal_dbi}) that $v_{\bm k}$ satisfies the following
equation of motion~\cite{Garriga:1999vw},
\begin{equation}
\frac{{\rm d}^2v_{\bm k}}{{\rm d}\eta ^2}
+\left(\frac{k^{2}}{\gamma^{2}}-
\frac{1}{z}\frac{{\rm d}^2 z}{{\rm d}\eta ^2}\right)v_{\bm k}
=0\,.
\label{eq:v}
\end{equation}
Here $\eta $ denotes the conformal time with which the scale factor is
expressed as $a(\eta)= -1/(H\eta )$ during exponential inflation.  In
this regime we find ${\rm d}^2z/{\rm d}\eta ^2=2/\eta ^2$, so that the
normalized solution of Eq.~(\ref{eq:v}) reads
\begin{equation}
v_{\bm k}(\eta )=\frac{1}{\sqrt{2kc_{\rm s}}}\lmk 1
-\frac{i}{kc_{\rm s}\eta }\rmk \e^{-ikc_{\rm s}\eta}\,,  
\label{eq:vsol}
\end{equation}
where we set $c_{\rm s}=1/\gamma$ (which corresponds to the
perturbations' sound speed) and made the usual adiabatic choice of
initial conditions. [Note that, as discussed in
Ref.~\cite{Lorenz:2008et}, this choice corresponds to somewhat more
restrictive conditions for scalar DBI perturbations evolving according
to Eq.~(\ref{eq:v}), as it would be the case for their standard counterparts.]
Thus $\vert\delta\phi_{\bm k}\vert ^2$ behaves as
\begin{equation}
\vert\delta\phi_{\bm k}\vert ^2
=\frac{1}{a^2\gamma^3}\vert v_{\bm k}\vert^2 
\rightarrow  \frac{H^2}{2k^3}
\end{equation}
in the long wavelength limit. Note that, unlike in the standard case,
in the DBI picture the boundary between long and short wavelength
regimes is not given by the Hubble radius but by the ``sound horizon''
$c_{\rm s}H^{-1}=(\gamma H)^{-1}$, as can be seen in the solution
(\ref{eq:vsol}).

\par

Now, the separation of the scalar field into long and short wavelength
components,
\begin{widetext}
\begin{equation}
\phi(\vec{x},t)=\varphi(t)+\int\frac{{\rm d}^3{\bm k}}{(2\pi)^{3/2}}\,
\Theta(k-\varepsilon a \gamma H)\,
\lkk a_{\bm k}\delta\phi_{\bm k}(t)\e^{-i\bm k\cdot \bm x}+
a_{\bm k}^\dag\delta\phi_{\bm k}^\ast(t)\e^{i\bm k\cdot \bm x} \rkk,
\label{eq:separation}
\end{equation}
is precisely the essence of the stochastic inflation approach, where
$\varepsilon$ in Eq.~(\ref{eq:separation}) is a small parameter (not
to be confused with the first slow-roll parameter), and $a_{\bm k}$
and $a_{\bm k}^\dag$ are annihilation and creation operators,
respectively.  Using this expression we find that the noise term in
Eq.~(\ref{eq:DBInoiseadded}) can be expressed as
\begin{equation}
\mathcal{C}\xi(t)=\varepsilon\,a(t)\gamma H^2\,
\int\frac{{\rm d}^3\bm k}{(2\pi)^{3/2}}\,\delta(k-\varepsilon a\gamma H)\,
\lkk a_{\bm k}\delta\phi_{\bm k}(t)\e^{-i\bm k\cdot \bm x}+
a_{\bm k}^\dag\delta\phi_{\bm k}^\ast(t)\e^{i\bm k\cdot \bm x} \rkk,
\end{equation}
with $\delta\phi_{\bm k}$ given by the solution of $v_{\bm k}$, see
Eq.~(\ref{eq:vsol}). Here we have used the condition that time
dependence of $c_{\rm s}=1/\gamma$ must be weak, $\vert \dot{c}_{\rm
  s}/c_{\rm s}\vert \ll H$ (which corresponds to $\delta_{1}\ll1$),
which is required to justify our analysis of quantum fluctuations
based on the mode function Eq.~(\ref{eq:vsol}). We then find the
correlation function of the stochastic noise is given by
\begin{equation}
\mathcal{C}^2\langle\xi(t)\xi(t')\rangle
=\left.\varepsilon^2\,a^2\gamma^2H^4\,\frac{4\pi k^2}{(2\pi)^3}\,
|\delta\phi_{\bm k}|^2\right|_{k=\varepsilon a\gamma H}\,
\frac{1}{\varepsilon a\gamma H^2}\,\delta(t-t')
=\frac{H^3}{4\pi^2}\delta(t-t').
\end{equation}
\end{widetext}
Thus we conclude that the analogue of the Langevin equation including
a noise term in the DBI case is
\begin{equation}
\dot{\varphi}=-\frac{2}{\kappa}\frac{H'}{\gamma}
+\frac{H^{3/2}}{2\pi}\,\xi(t)\, .
\label{eq:DBILangevin}
\end{equation}
One notices that the factor $\gamma $ only appears in the classical
term and not in the normalization of the noise term.

\par

By means of this equation, one can estimate in which regime the
quantum effects are important. If the field's behavior is dominated by
quantum effects, one can neglect the classical drift in
Eq.~(\ref{eq:DBILangevin}). As already mentioned, using the properties
(\ref{eq:noise}) this leads to $\mean{\varphi ^2(t)}=H^3
t/(4\pi^2)$. However, the typical time scale $\Delta t$ now is
$1/(H\gamma)$, since the ``horizon'' felt by the scalar field (and its
perturbations) shrinks by a factor of $1/\gamma$ compared to the
standard case. Therefore, the typical quantum kick undergone by the
field in the characteristic time scale is %jy
\begin{equation}
\Delta\phi_{\rm qu}=\sqrt{\mean{\varphi ^2(t)}}
=\sqrt{\frac{H^3}{4\pi ^2}\frac{1}{\gamma H}}
=\frac{H}{2\pi\gamma ^{1/2}}\, .
\end{equation}
On the other hand, if the quantum effects are negligible, then the
equation determining the behavior of the field is nothing but the
slow-roll equation of motion and this implies that
\begin{equation}
\Delta\phi_{\rm cl}=-\frac{V'}{3H\gamma }\Delta t
=-\frac{V'}{3H^2\gamma ^2}\, .
\end{equation}
Setting $\Delta\phi_{\rm qu}= \Delta\phi_{\rm cl}$, one easily
concludes that the corresponding vacuum expectation value (vev) of the
field $\phi_{*}$ obeys the equation
\begin{equation}\label{eq:defquantum}
H(\phi_{*})=\frac{\mpl^2}{4}\frac{V'(\phi_{*})}{V(\phi_{*})}
\frac{1}{\left[\gamma(\phi_{*})\right]^{3/2}}\, .
\end{equation}
(Again, in the standard case $\gamma =1$ and one recovers the usual
criterion.) Eq.~(\ref{eq:defquantum}) allows us to decide for which
values of the inflaton quantum effects play an important r\^{o}le.

\par

So far we have used the cosmic time $t$ as the time variable in the
Langevin equation, see Eq.~(\ref{eq:DBILangevin}).  The choice of the
time variable, however, is a subtle issue as different choices may
lead to physically inequivalent results. It has recently been
advocated in Refs.~\cite{Finelli:2008zg,Finelli:2010sh} that in many
cases it would be more appropriate to use the number of e-folds,
$N=\ln a$, as the time variable depending on what quantities we wish
to calculate. Indeed, in this case, the results obtained from the
stochastic formalism coincide with those derived from perturbative
quantum field theory~\cite{Finelli:2008zg,Finelli:2010sh}. Written in
terms of the number of e-folds, the Langevin equation
(\ref{eq:DBILangevin}) reads
\begin{equation}
\label{eq:langevinefold}
\frac{{\rm d}\varphi}{{\rm d}N}=-\frac{2}{\kappa \gamma}\frac{H'}{H}
+\frac{H}{2\pi}\xi(N)\, ,
\end{equation}
where $\xi (N)$ is a new stochastic process (for which, 
allowing for slightly slippery notation, 
we still denote by the same symbol) such that
$\mean{\xi(N)\xi(N')}=\delta(N-N')$. It is easy to check that a free
field satisfies $\mean{\varphi^2(N)}=H^2(N-N_\ini)/(4\pi^2)$ as
expected. Below, we will carry out our calculations in both time variables,
cosmic time $t$ and the number of e-folds $N$.

\subsection{Solving the DBI Langevin Equation}
\label{subsec:solving}

We now solve Eq.~(\ref{eq:DBILangevin}) using a perturbative expansion
in the noise as shown in Ref.~\cite{Martin:2005ir}. We use the following
ansatz for the Hubble-patch averaged field,
\begin{equation}
\label{eq:ansatz}
\varphi(t)=\varphi_\cl(t)+\delta\varphi_{1}(t)+\delta\varphi_{2}(t)+\dots,
\end{equation}
where $\varphi_\cl(t)$ is the classical field,
$\delta\varphi_{1}(t)\propto\mathcal{O}(\xi)$ and
$\delta\varphi_{2}(t)\propto\mathcal{O}(\xi^{2})$. In principle, this
expansion can be carried to any order in $\xi$. At zeroth order, we get
back the classical slow-roll equation~(\ref{eq:KG-DBIsr}). At first
order, one obtains an equation for $\delta\varphi_{1}(t)$, namely
\begin{align}
\frac{\dd\delta\varphi_{1}(t)}{\dd t}
& +\frac{2}{\kappa}\frac{H'(\varphi_\cl)}{\gamma(\varphi_\cl)}
\left[\frac{H''(\varphi_\cl)}{H'(\varphi_\cl)}
-\frac{\gamma'(\varphi_\cl)}{\gamma(\varphi_\cl)}\right]
\delta\varphi_{1}(t)\nonumber \\ 
&= \frac{H^{3/2}(\varphi_\cl)}{2\pi}\,\xi(t)\, .
\end{align}
As expected, this equation differs from the standard one by the presence
of the Lorentz factor and its derivative. At second order in the noise,
one obtains the equation describing the evolution of
$\delta\varphi_{2}(t)$,
\begin{align}
\frac{\dd\delta\varphi_{2}(t)}{\dd t}
& +\frac{2}{\kappa}\frac{H'(\varphi_\cl)}{\gamma(\varphi_\cl)}
\left[\frac{H''(\varphi_\cl)}{H'(\varphi_\cl)}
-\frac{\gamma'(\varphi_\cl)}{\gamma(\varphi_\cl)}\right]\delta\varphi_{2}(t)
\nonumber \\
&= \frac{3}{4\pi}\,H'(\varphi_\cl)\,H^{1/2}(\varphi_\cl)\,\xi(t)\,
\delta\varphi_{1}(t) \nonumber \\
& -\frac{1}{\kappa}\,
\frac{H'(\varphi_\cl)}{\gamma(\varphi_\cl)}
\biggl[\frac{H'''(\varphi_\cl)}{H'(\varphi_\cl)}
-\frac{\gamma''(\varphi_\cl)}{\gamma(\varphi_\cl)}
+2\frac{\gamma'^{2}(\varphi_\cl)}{\gamma^{2}(\varphi_\cl)}
\nonumber \\
& -2\frac{H''(\varphi_\cl)}{H'(\varphi_\cl)}
\frac{\gamma'(\varphi_\cl)}{\gamma(\varphi_\cl)}\biggr]
\delta\varphi_{1}^{2}(t)\, .
\end{align}
The same remark as before is valid: the equation for
$\delta\varphi_{2}(t)$ is modified by the presence of the factor
$\gamma$ and its derivatives. Clearly, the equation contains
derivatives of $\gamma $ up to second order because it is second order
in the noise expansion.

\par

We are now in a position to solve the above equations. Since they are
first order differential equations, they can be solved by varying the
integration constant. One finds for $\delta\varphi_{1}(t)$ that
\begin{widetext}
\begin{eqnarray}
\delta\varphi_{1}(t)&=&
\frac{H'[\varphi_\cl(t)]}{2\pi\gamma[\varphi_\cl(t)]}
\int_{t_{\mathrm{in}}}^{t}\dd
 t'\,\frac{H^{3/2}[\varphi_\cl(t')]}{H'[\varphi_\cl(t')]}\,
\gamma[\varphi_\cl(t')]\,\xi(t')\, ,\nonumber \\
\end{eqnarray}
while for $\delta\varphi_{2}(t)$ one obtains
\begin{eqnarray}
\delta\varphi_{2}(t)&=& \frac{3}{4\pi}\frac{H'[\varphi_\cl(t)]}
{\gamma[\varphi_\cl(t)]}\int_{t_{\mathrm{in}}}^{t}\dd t'\,
H^{1/2}[\varphi_\cl(t')]\,\gamma[\varphi_\cl(t')]\,\xi(t')\,
\delta\varphi_{1}(t')\nonumber \\
& &
-\frac{H'[\varphi_\cl(t)]}{\kappa\gamma[\varphi_\cl(t)]}
\int_{t_{\mathrm{in}}}^{t}\dd t'
\biggl\{\frac{H'''[\varphi_\cl(t')]}{H'[\varphi_\cl(t')]}
-\frac{\gamma''[\varphi_\cl(t')]}{\gamma[\varphi_\cl(t')]}
 +2\frac{\gamma'^{2}[\varphi_\cl(t')]}{\gamma^{2}[\varphi_\cl(t')]}
-2\frac{H''[\varphi_\cl(t')]}{H'[\varphi_\cl(t')]}
\frac{\gamma'[\varphi_\cl(t')]}{\gamma[\varphi_\cl(t')]}\biggr\}
\delta\varphi_{1}^{2}(t')\, .
\end{eqnarray}
\end{widetext}

\par
From the above expressions and the properties of the noise given by
Eq.~(\ref{eq:noise}), it is obvious that
\(\mean{\delta\varphi_{1}}=0\), and for the second moment we find
\begin{equation}
\label{eq:dbidf1}
\mean{\delta\varphi_{1}^{2}}
=\frac{\kappa}{2}\left(\frac{H'}{2\pi\gamma}\right)^{2}
\int_{\varphi_{\mathrm{cl}}}^{\varphi_{\mathrm{in}}}\dd\psi\,
\left[\frac{H(\psi)\gamma(\psi)}{H'(\psi)}\right]^{3}\, .
\end{equation}
The next step is to calculate $\mean{\delta\varphi_{2}}$. Lengthy but
straightforward calculations lead to
\begin{align}
\mean{\delta\varphi_{2}}=&
\frac{H'}{2\pi\mpl^{2}\gamma}\Biggl\{\left(\frac{H'}{\gamma}\right)'
\int_{\varphi_\cl}^{\varphi_{\mathrm{in}}}\dd\psi\,
\left(\frac{H\gamma}{H'}\right)^{3}
\nonumber \\
&-\int_{\varphi_\cl}^{\varphi_{\mathrm{in}}}\dd\psi\,
\left[\left(\frac{H\gamma}{H'}\right)^{3}
\left(\frac{H'}{\gamma}\right)'
-\frac{3}{2}\frac{H^{2}\gamma^{2}}{H'}\right]\Biggr\}\nonumber\\
&= \frac{\left(H'/\gamma\right)'}{2\left(H'/\gamma\right)}
\mean{\delta\varphi_{1}^{2}}
\nonumber \\
&+\frac{H'/\gamma}{4\pi\mpl^{2}}
\left[\left(\frac{\gamma^{2}H^{3}}{H'^{2}}\right)_{\varphi_{\mathrm{in}}}
-\left(\frac{\gamma^{2}H^{3}}{H'^{2}}\right)_{\varphi_\cl}\right]\, .
\label{eq:dbidf2}
\end{align}
As in the standard case, everything can be reduced to the calculation
of a single quadrature. As expected in the DBI case, this quadrature
contains the factor $\gamma$.

\par

Using these results, one can now calculate the %jy%
PDF, $P_{\rm c}(\varphi,t)$, which describes the probability of the
stochastic process $\varphi[\xi]$ to take %jy%
a given value $\varphi$ at a given time $t$ in a single coarse-grained
domain (see Ref.~\cite{Martin:2005ir}),
\begin{align}
\label{eq:Pc} 
P_{\rm c}(\varphi,t) &=\mean{\delta(\varphi-\varphi[\xi])}\nonumber \\
& =\frac{1}{\sqrt{2\pi\mean{\df_1^2}}}\,
\exp\left[-\frac{(\varphi-\varphi_\cl-\mean{\df_2})^{2}}
{2\mean{\df_1^2}}\right]\nonumber\\
&\equiv P_{\rm g}(\varphi-\varphi_\cl-\mean{\df_2})\, .
\end{align} [In the last line, we have introduced the definition
$P_{\rm g}$ for later use, see Eq.~(\ref{eq:Pg}).] If, however, one is
interested in spatial averaging over the entire Universe (instead of a
single domain), one has to include a weight factor
$a^{3}(\varphi)=\exp\left[3\int\dd\tau\,H(\varphi)\right]$ for the
physical volume of each Hubble-sized domain. This leads to
\begin{align}
\label{eq:volume}
P_{\rm v}(\varphi,t) &=
\frac{\mean{\delta(\varphi-\varphi[\xi])\,\e^{3\int\dd\tau
H(\varphi[\xi])}}} {\mean{\e^{3\int\dd\tau H(\varphi[\xi])}}}
\nonumber \\
&=\frac{1}{\sqrt{2\pi\mean{\df_1^2}}}
\exp\left[ -\frac{\left(\varphi-\mean{\varphi}
-3I^{\rm T}J\right)^2}{ 2\mean{\df_1^2}}\right] \nonumber \\
& =  P_{\rm g}(\varphi-\mean{\varphi}-3I^{\rm T}J)
\, ,
\end{align}
where again the last line is a definition used in later Sections. In
Eq.~(\ref{eq:volume}), $\mean{\varphi}$ is the single domain averaged
mean value $\mean{\varphi}\!=\varphi_\cl+\mean{\df_2}$, to which the
volume effects induce the additional correction given by
\begin{eqnarray}
\label{eq:voleffects}
3I^{\rm T}J &=&
3\int^{t}_{t_{\ini}}\dd\tau\,H'(\tau)\mean{\df_{1}(t)
\df_{1}(\tau)}\nonumber\\
&=&\frac{12H'}{\mpl^{4}\gamma}
\int_{\varphi_\cl}^{\varphi_{\ini}}\dd\psi\,
\frac{H^{4}\gamma^{3}}{H'^{3}}
-\frac{12\pi\gamma H}{\mpl^{2}H'}\mean{\delta\varphi_{1}^{2}}.
\label{eq:3ITJ}
\end{eqnarray}
Once this integration carried out, the volume-weighted distribution
$P_{\rm v}(\varphi,t)$ as defined in Eq.~(\ref{eq:volume}) is also
completely determined. Again, comparing the results of this section
with those for the standard case presented in
Refs.~\cite{Martin:2005ir,Martin:2005hb}, we see that the only changes
are the additional powers of $\gamma$ found in $\df_{1}(t)$ and
$\df_{2}(t)$. Below, we calculate $P_{\rm c}(\varphi,t)$ and $P_{\rm
  v}(\varphi,t)$ for an exemplary shape of $V(\varphi)$ and
$T(\varphi)$.

\par

As discussed above, the Langevin equation can also be written with the
number of e-folds as the time variable. It is straightforward to
repeat the above analysis for the corresponding Langevin equation
(\ref{eq:langevinefold}). In particular, the first and second order
corrections obtained in terms of e-folds read
\begin{equation}
\df_1(N)=\frac{1}{2\pi}\frac{H'}{H\gamma }\int _{N_\ini}^N{\rm d}N'\,
\frac{H^2(N')\,\gamma(N')}{H'(N')}\,\xi(N'),
\end{equation}
and 
\begin{eqnarray}
\df_2(N) &=& \frac{1}{2\pi}\frac{H'}{H\gamma }
\int _{N_\ini}^N{\rm d}N'\,H\gamma\, 
\df_1(N')\,\xi(N') \label{eq:dphi2efolds}\\  
&-&\frac{1}{\kappa }\frac{H'}{H\gamma }
\int _{N_\ini}^N{\rm d}N'\frac{H\gamma }{H'}
\left(\frac{H'}{H\gamma }\right)''
\df_1^2(N').\nonumber 
\end{eqnarray}
In Eq.~(\ref{eq:dphi2efolds}), arguments in the integrands 
have been partially suppressed where they are evident. 
As before, it is easy to calculate $\mean{\df_1^2}$
and $\mean{\df_2}$ from these expressions. It is found that
\begin{equation}
\label{eq:meandf1efold}
\mean{\df_1^2}_{N}=\frac{\kappa }{8\pi^2}\frac{H'^2}{H^2\gamma ^2}
\int^{\varphi_{\ini}}_{\varphi_\cl}\frac{H^5\gamma ^3}{H'^3}{\rm
d}\psi\, .
\end{equation}
For the case of a standard kinetic term (hence, $\gamma =1$), this
equation coincides with Eq.~(40) of
Ref.~\cite{Finelli:2010sh}. Furthermore, we find from
Eq.~(\ref{eq:dphi2efolds}) that
\begin{eqnarray}
\label{eq:meandf2efold}
\mean{\df_2}_{N}&=&\frac{H\gamma }{2H'}\left(\frac{H'}{H\gamma }\right)'
\mean{\df_1^2}+\frac{\kappa }{32\pi^2}\frac{H'}{H\gamma }
\biggl[\left(\frac{H^4\gamma ^2}{H'^2}\right)_\ini
\nonumber \\ & &
-\left(\frac{H^4\gamma ^2}{H'^2}\right)\biggr].
\end{eqnarray}
Again, if $\gamma =1$, this equation corresponds to Eq.~(48) of
Ref.~\cite{Finelli:2010sh}. Let us notice that, as in our previous
results of Eqs.~(\ref{eq:dbidf1}) and (\ref{eq:dbidf2}), the
expression (\ref{eq:meandf1efold}) for $\mean{\df_1^2}_{N}$ is
sufficient to obtain $\mean{\df_2}_{N}$ in
Eq.~(\ref{eq:meandf2efold}), \ie no additional quadrature is
necessary.

\par

Following the same steps, one can also evaluate the PDFs in terms of
e-folds. The corresponding probability $P_{\rm c}^{(N)}(\varphi, t)$
is similar to Eq.~(\ref{eq:Pc}), except that now the
formulas~(\ref{eq:meandf1efold}) and~(\ref{eq:meandf2efold}) should be
used in their respective places.  The definition of the volume
weighted distribution also remains the same, see
Eq.~(\ref{eq:volume}), but the term $3I^{\rm T}J$ now reads
\begin{eqnarray}
\label{eq:volumeefold}
\left(3I^{\rm T}J\right)_{N} &=& 
\frac{12}{\mpl^4}\frac{H'}{H\gamma }
\int _{\varphi_\cl}^{\varphi_\ini}{\rm d}\psi\, \frac{H^5\gamma ^3}{H'^3}\,
\ln\left(\frac{H}{\mpl}\right)\nonumber \\ & &
-\frac{12\pi}{\mpl^2}\frac{H\gamma }{H'}\mean{\df_1^2}\,
\ln\left(\frac{H}{\mpl}\right)
%
%\left(3I^{\rm T}J\right)_{N} &=& 
%\frac{12\pi}{\mpl^4}\frac{H'}{H\gamma }
%\int _{\varphi_\cl}^{\varphi_\ini}{\rm d}\psi \frac{H^5\gamma ^3}{H'^3}
%\ln\left(\frac{H}{\mpl}\right)\nonumber \\ & &
%-\frac{12\pi}{\mpl^2}\frac{H\gamma }{H'}
%\mean{\df_1^2}\ln\left(\frac{\varphi_\cl}{\varphi_\ini}\right)\,.
\end{eqnarray} 
This equation should be compared to Eq.~(\ref{eq:3ITJ}).  Again,
assessment of the volume effects requires the calculation of a new
quadrature.

\section{Application to Brane Inflation}
\label{sec:applications}

In the following, we apply the formalism developed in
Sec.~\ref{sec:DBI} to a popular class of string-inspired inflation
models with DBI kinetic term. We focus on scenarios of the brane
inflation type, for which the inflaton field corresponds to the
position of a $D3$-brane embedded in a higher-dimensional
background. Successful model building requires that the six extra
dimensions be deformed in a way described by a warp factor $T(\phi)$;
the resulting geometry is commonly called a Klebanov Strassler
throat. Scenarios of this type have been the subject of a vast body of
literature, see \eg
Refs.~\cite{Dvali:1998pa,Burgess:2001fx,Kachru:2003sx,Baumann:2006th,Baumann:2007np,Baumann:2007ah,
  Krause:2007jk,Pajer:2008uy,Baumann:2008kq,Chen:2004gc,Chen:2005ad,Bean:2006qz,
  Chen:2006hs,Bean:2007eh}. For our purposes, we denote the warp
factor and the inflationary potential by
\begin{equation}
\label{eq:pot-const}
T(\phi)=\frac{\phi^{4}}{\lambda}\, ,\quad
V(\phi)=V_{0}\left[1-\left(\frac{\mu}{\phi}\right)^4\right]
+ \dfrac{\varepsilon}{2}\,m^{2}\phi^{2}\, ,
\end{equation}
where $\varepsilon =\pm 1$. The plus sign identifies so-called
``Ultra-Violet'' (UV) models (where the $D3$ moves from the edge
towards the bottom of the throat geometry and hence the field value
decreases during inflation), while the minus sign refers to the
``Infra-Red'' (IR) setting (where the $D3$ climbs out of the throat
and the inflaton's field value grows with time).

\par

For completeness, the formulation Eq.~(\ref{eq:pot-const}) of
$V(\phi)$ includes a Coulomb potential term due to the $D3$'s
attraction towards a $\bar{D}3$-brane sitting at the bottom of the
throat. Originally, the small Coulombic attraction resulting from this
very flat ($\propto 1/\phi ^4$) potential was considered the
inflaton's only driving force, ignoring the (potentially much steeper)
second term $\propto\phi^{2}$ in Eq.~(\ref{eq:pot-const})
\cite{Kachru:2003sx}.  In fact, it was shown that in this case the DBI
dynamics do not affect the inflationary evolution
\cite{Lorenz:2007ze}, and the stochastic effects were also assessed in
the same reference.  Therefore, in the following, we ignore the
$\propto1/\phi ^4$ Coulombic contribution in
Eq.~(\ref{eq:pot-const}). Note, however, that conceptually the
presence of an anti-brane at the bottom of the KS throat shall be
important for our reasoning in later Sections. The quadratic potential term in
Eq.~(\ref{eq:pot-const}) has a different status. We shall treat it
here as a phenomenological description of the potential that the
various background moduli fields produce for the mobile
$D3$-brane. The exact shape of these moduli contributions is still a
subject of active
research~\cite{Kachru:2003sx,Baumann:2006th,Baumann:2007np,Baumann:2007ah,
  Krause:2007jk,Pajer:2008uy,Baumann:2008kq}.

\par

From the cosmological point of view, the inflation model of
Eq.~(\ref{eq:pot-const}) has three parameters, the mass $m$, the
dimensionless constant $\lambda $ and $V_0$ (with dimension
$\mpl^{4}$). In fact, as we show below, it is rather two dimensionless
combinations of $(m,\lambda,V_{0})$, which we shall call
$(\alpha,\beta)$, that characterize the evolution. We define these
parameters by
\begin{equation}
\label{eq:alpha}
\alpha \equiv \frac{12\pi \mpl ^2}{\lambda m^2}
=\frac{96\pi ^2}{\kappa \lambda m^2}\, , \qquad 
\beta \equiv \dfrac{V_0}{m^2\mpl^2}\, .
\end{equation}
Physically, $\beta$ measures the importance of the constant term
relatively to the mass term in the potential (recall that we are
neglecting the Coulomb term which involves the parameter $\mu$).

\par

The geometric interpretation (in terms of an extra-dimensional brane
position) of the inflaton field enforces some intrinsic consistency
conditions. The renormalization relating the inflaton field $\phi$ to
the (radial) throat coordinate $r$ reads $\phi =\sqrt{T_3}r$, where
$T_3=[(2\pi)^3g_{_{\rm S}}\alpha '^2]^{-1}$ is the tension of the
$D3$-brane, calculated from the string coupling $g_{_{\rm S}}$ and
scale $\alpha '$. (We do not consider a possible motion of the brane
along angular coordinates of the throat.) Let the edge of the throat
correspond to some $r_{\UV}$, where the KS corner of the geometry is
connected to the compactified six-dimensional bulk. Since the metric
outside the throat is unknown, one has to impose $r<r_{\UV}$ to ensure
that the brane stays inside the well-defined KS region. In terms of
stringy background parameters, one can express
\begin{equation}\label{eq:rUV}
r_{\UV}^4=4\pi g_{_{\rm S}}\alpha '^2\frac{{\cal N}}{v}\, ,
\end{equation}
where ${\cal N}$ is a positive integer representing the total
Ramond-Ramond (RR) charge and $v$ represents the (dimensionless)
parameter measuring the volume of the five-dimensional submanifold
that forms the basis of the 6d throat in units of the five-sphere
volume. Via the inflaton's renormalization, this evidently is an upper
bound on the inflaton $\phi$. Note that, depending on whether we are
talking about UV or IR models, this constraint affects the initial or
final field value, $\phi_{\ini}^{(\UV)}<\phi_{\UV}$ or $\phi_{\rm
  end}^{(\IR)}<\phi_{\UV}$.

\par

A second consistency condition is the requirement that the volume of the
throat to be smaller than the total volume of the compactified extra
dimensions. This total volume has observational significance since it
enters into the four-dimensional Planck mass $\mpl$. From this
constraint it follows that
\begin{equation}\label{eq:volumeconstraint}
\phi <\phi_{\UV}<\frac{\mpl}{\sqrt{2\pi {\cal N}}}\, ,
\end{equation}
This means that inflation always occurs for sub-Planckian values of
$\phi$. On the other hand, the bottom of the throat being located at
$r_0$, one must have $\phi>\phi_0\equiv \sqrt{T_3}r_0$.  Moreover, for
the model to be valid, the (physical) distance between the brane must
be larger than the string length and one can show that this amounts to
\begin{equation}
\phi >\phi_{\rm strg}=\phi _0
{\rm e}^{\sqrt{\alpha'}{r_{\UV}}}\, .
\end{equation}

\par

Note also that the parameters of the warp factor and potential in
Eqs.~(\ref{eq:pot-const}) can be calculated in terms of the stringy
parameters. Physically, $T(\phi)$ is the position-dependent brane
tension and it can be written as $T(\phi)=T_3\left(\phi/\phi
  _{\UV}\right)^4$, which implies [compare Eqs.~(\ref{eq:pot-const})
and (\ref{eq:rUV})] that
\begin{equation}
\lambda =\frac{{\cal N}}{2\pi ^2v}\, .
\end{equation}
The constant term $V_0$ is given by $V_0=4\pi^2v\phi_0^4/{\cal N}$,
which can also be expressed as
\begin{equation}
V_0=2\,h^4(r_0)\,T_3\, ,
\end{equation}
where $h(\phi)\equiv \phi /\phi_{\UV}$ is the warping function as
it appears in the 10d metric [it holds that $T(\phi)\propto
h^{4}(\phi)$].

\par

It turns out that Eq.~(\ref{eq:volumeconstraint}) can be rewritten in a
 more quantitative way.  Since the volume of a Klebanov Strassler throat
 is known, $V_6^{\rm throat}=2\pi^4g_{_{\rm S}}{\cal N}\alpha
 '^2r_{\UV}^2$, and the total six-dimensional volume is related to the
 four-dimensional Planck mass, $V_6^{\rm tot}=\mpl^2(2\pi)^7g_{_{\rm
 S}}^2\alpha '^4/(16\pi)$, one deduces for our parameters
 $(\alpha,\beta)$ defined in Eq.~(\ref{eq:alpha}) that
\begin{equation}
\label{eq:conditionba}
\sqrt{\frac{\beta
 }{\alpha}}<\frac{1}{\sqrt{24\pi^3}}\frac{h^2(r_0)}{{\cal N}}\ll 1\, .
\end{equation}
A similar equation [see Eq.~(2.10)] was used in Ref.~\cite{Bean:2007eh}.

\subsection{Chaotic Klebanov Strassler Inflation}
\label{subsec:chaotic} 

In our first example, we set the parameter $\beta=0$, hence we are
considering the case of ``pure'' CKS inflation without a constant
term.  The potential and warp factor hence are given by
\begin{equation}
\label{eq:chaotic}
V(\phi)=\frac{1}{2}\,m^{2}\phi^{2},\qquad
T(\phi)=\frac{\phi^{4}}{\lambda}\, .  
\end{equation}
In the slow-roll limit, it follows from Eq.~(\ref{eq:Hdbi}) that
$H^{2}\simeq4\pi m^{2}\phi^{2}/(3\mpl^{2})$ (see Appendix
\ref{app:srlimit}) and, therefore, the Lorentz factor calculated from
Eq.~(\ref{eq:gamma}) behaves as
\begin{equation}
\label{eq:gammaCKS}
\gamma(\phi)=\sqrt{1+\frac{2\lambda
m^{2}}{3\kappa}\frac{1}{\phi^{4}}}
=\frac{\mpl^{2}}{\phi^{2}}\sqrt{\frac{\phi^{4}}{\mpl^{4}}
+\frac{1}{\alpha}}\, .
\end{equation}
If we plug these expressions into Eq.~(\ref{eq:dbidf1}), the
calculation of $\mean{\df_{1}^{2}}$ can be reduced to an exactly
solvable integral, and the final result for $\mean{\df_{1}^{2}}$ is
\begin{eqnarray}
\mean{\df_{1}^{2}}&=&\frac{4m^{2}}{3\gamma^{2}}
\bigg[\frac{1}{2\alpha}\left(\gamma-\gamma_\ini\right)
-\frac{1}{4}\left(\frac{\varphi_\cl^{4}}{\mpl^4}
\gamma-\frac{\varphi_\ini^{4}}{\mpl^4}\gamma_\ini\right)
\nonumber\\
& &+\frac{3}{2\alpha}\,\ln\left(\frac{\varphi_{\ini}}{\varphi_\cl}\right)
+\frac{3}{4\alpha}\,\ln\left(\frac{1+\gamma_\ini}
{1+\gamma}\right)\bigg]\, ,
\end{eqnarray} 
where $\gamma=\gamma(\varphi_\cl)$ and
$\gamma_\ini=\gamma(\varphi_\ini)$. For $\mean{\df_{2}}$ we have from
Eq.~(\ref{eq:dbidf2}) that
\begin{equation}
\mean{\df_{2}}=\frac{\gamma^{2}-1}
{\gamma^{2}}\frac{1}{\varphi_\cl}
\mean{\df_{1}^{2}}+\frac{m^{2}}{3\gamma\mpl^{4}}
\left(\gamma_\ini^{2}\varphi_\ini^{3}
-\gamma^{2}\varphi_\cl^{3}\right)\, .
\end{equation}
To calculate the volume effects, another integration is
necessary. With the potential and warp factor given by
Eq.~(\ref{eq:chaotic}), one finds from Eq.~(\ref{eq:3ITJ}) that
\begin{widetext}
\begin{eqnarray}
3I^{T}J &=& \frac{16\pi}{\gamma}\frac{m^{2}}{\mpl^{2}}\,
\mpl \Biggl\{-\frac{\mpl}{\varphi _\ini}\left(
\frac{\varphi_\ini^4}{\mpl^4}+\frac{1}{\alpha}\right)^{3/2}
+\frac{\mpl}{\varphi _\cl}\left(
\frac{\varphi_\cl^4}{\mpl^4}+\frac{1}{\alpha}\right)^{3/2}
-\frac32 (-1)^{-1/4}\alpha ^{-5/4}\Biggl[
B\left(-{\alpha}\frac{\varphi_\ini^{4}}{\mpl^{4}},
\frac34,\frac32\right)\nonumber \\
& & -B\left(-{\alpha}\frac{\varphi_\cl^{4}}{\mpl^{4}},
\frac34,\frac32\right)\Biggr]
\Biggr\}-12\pi\gamma\,\frac{\mean{\df_{1}^{2}}}
{\mpl^{2}}\,\varphi_{\cl}\, ,
\end{eqnarray}
\end{widetext}
where $B$ is the incomplete Euler's integral of the first kind defined
by $B(z,a,b)\equiv \int _0^zt^{a-1}(1-t)^{b-1}{\rm
d}t$~\cite{Abramovitz:1970aa,Gradshteyn:1965aa}. Notice that, in the
above equation, the function $B$ takes in fact complex values. However,
multiplied by the factor $(-1)^{-1/4}$, the result is real as it should
be.

\par

Let us discuss these results in more detail in the light of the
consistency constraint Eq.~(\ref{eq:volumeconstraint}). In the limit
of small field values compared to the Planck mass, the Lorentz factor
is large and can be approximated by $\gamma (\varphi_\cl)\simeq
\mpl^2/\left(\sqrt{\alpha}\varphi_\cl^2\right)$. It is hence easy to
show that
\begin{eqnarray}
\label{eq:meandf1approx}
\mean{\df_{1}^{2}}&\simeq & \frac{2m^2}{3\sqrt{\alpha}}
\left(\frac{\varphi_\cl}{\mpl}\right)^2, \\
\label{eq:meandf2approx}
\mean{\df_{2}}&\simeq & \frac{m^2}{3\mpl^2}
\frac{\varphi_\cl}{\sqrt{\alpha}} \, .
\end{eqnarray}
The expression giving the volume effects can also be simplified.  Notice
 that $B(z,a,b)=z^a{}_2F_1(a,1-b,a+1,z)$, where ${}_2F_1$ is the
 hypergeometric function~\cite{Abramovitz:1970aa,Gradshteyn:1965aa}.
 Using the asymptotic behavior of
 ${}_2F_1$~\cite{Abramovitz:1970aa,Gradshteyn:1965aa}, straightforward
 calculations show that
\begin{equation}
\label{eq:volumeapprox}
3I^{T}J \simeq 8\pi \frac{m^2}{\mpl^2}\frac{ \varphi_\cl}{\alpha} \, .
\end{equation}

Let us now establish the results in terms of the numbers of
e-folds. Using Eq.~(\ref{eq:meandf1efold}), one obtains
\begin{eqnarray}
\mean{\df_1^2}_{N} &=& -\frac{2}{3}\frac{m^2}{\gamma ^2}
\frac{\mpl^2}{\phi^2}\biggl[
\frac{\gamma ^3}{3}\frac{\varphi_\cl^6}{\mpl^6}
-\frac{\gamma ^3_\ini}{3}\frac{\varphi_\ini^6}{\mpl^6}
+\frac{\gamma }{\alpha}\frac{\varphi_\cl^2}{\mpl^2}
\nonumber \\ & &
-\frac{\gamma _\ini}{\alpha}\frac{\varphi_\ini^2}{\mpl^2}
-\frac{1}{\alpha ^{3/2}}\,{\rm arcsinh} \left(
\frac{1}{\alpha ^{1/2}}\frac{\mpl^2}{\varphi^2_\cl}\right)
\nonumber \\ & &
+\frac{1}{\alpha ^{3/2}}\,{\rm arcsinh} \left(
\frac{1}{\alpha ^{1/2}}\frac{\mpl^2}{\varphi^2_\ini}\right)
\biggr].
\end{eqnarray}
One can easily check that this expression vanishes at initial time as
expected. The corresponding expression for $\mean{\df_2}$ can be
deduced from Eq.~(\ref{eq:meandf2efold}). The result reads
\begin{eqnarray}
\mean{\df_2}_{N} &=& \frac{1}{2\varphi_\cl}\frac{\gamma ^2-2}{\gamma ^2}
\mean{\df_1^2}\nonumber \\ & & 
+\frac{1}{3\gamma}\frac{m^2}{\varphi_\cl}
\left(\gamma_\ini^2\frac{\varphi_\ini^4}{\mpl^4}-
\gamma ^2\frac{\varphi ^4_\cl}{\mpl^4}\right)
\end{eqnarray}
In the limit where the Lorentz factor $\gamma $ is large, the above
expressions can be approximated by
\begin{eqnarray}
\mean{\df_1^2}_{N}&\simeq& \frac{2m^2}{3\sqrt{\alpha }}\left(
\frac{\varphi_\cl}{\mpl}\right)^2
\ln \left(\frac{\varphi_\ini}{\varphi_\cl}\right)^{2}, \\
\mean{\df_2}_{N} &\simeq & \frac{2m^2}{3\mpl^2}
\frac{\varphi_\cl }{\sqrt{\alpha}}
\ln\left(\frac{\varphi_\ini}{\varphi_\cl}\right).
\end{eqnarray}
It is interesting to compare these formulas to
Eqs.~(\ref{eq:meandf1approx}) and~(\ref{eq:meandf2approx}). We see
that working in terms of the number of e-folds simply introduces
(apart from numerical prefactors) a logarithmic correction to the
correlation functions: roughly speaking, the new correlation functions
are obtained from the old ones with the replacement $\varphi_\cl
\rightarrow \varphi_\cl\ln(\varphi_\cl/\varphi_\ini)$. This is
confirmed by a calculation of the volume effect. Using
Eq.~(\ref{eq:volumeefold}), one obtains
\begin{eqnarray}
\left(3I^{\rm T}J\right)_{N} &\simeq &\frac{16\pi}{\alpha} 
\frac{m^2}{\mpl^2}\,\varphi_\cl\nonumber\\&&
\times\Biggl\{\ln\left(\frac{\varphi_\cl}{\varphi_\ini}\right)\,
\ln\left[\left(\frac{4\pi}{3}\right)^{1/2}\frac{m\varphi_\cl}{\mpl^2}\right]
\nonumber \\ & & 
+\frac12\ln ^2\left[\left(\frac{4\pi}{3}\right)^{1/2}
\frac{m\varphi_\ini}{\mpl^{2}}\right]
\nonumber \\ & &
-\frac12\ln ^2\left[\left(\frac{4\pi}{3}\right)^{1/2}
\frac{m\varphi_\cl}{\mpl^{2}}\right]\Biggr\}.
\end{eqnarray}
This formula should be compared to Eq.~(\ref{eq:volumeapprox}).

\par

In order to see whether the stochastic effects are important or not,
we must normalize the model's parameters to the COBE
observations. This was done in Ref.~\cite{Lorenz:2008et}, where it was
shown that [see that reference's Eq.~(127)]
\begin{equation}
\left(\frac{m}{\mpl}\right)^2
=\frac{45}{4\pi}\frac{Q^2}{T_{_{\rm CMB}}^2}\frac{1}{\alpha}\, ,
\end{equation}
where the quantity $Q^2/T_{_{\rm CMB}}^2$ can be expressed in terms of
the CMB quadrupole according to
\begin{equation}
\frac{Q}{T_{_{\rm CMB}}}=\sqrt{\frac{5C_2}{4\pi}}\simeq 6\times
 10^{-6}\, .
\end{equation}
In Ref.~\cite{Lorenz:2008et} [see Eq.~(116)], it was also demonstrated
that the first slow-roll parameter $\epsilon _1 $ (see
Appendix~\ref{app:srlimit} for a precise definition of the slow-roll
hierarchy), for the model under consideration, can be expressed as
$\epsilon _1\simeq \sqrt{\alpha}/(4\pi)$. Therefore, in a realistic
inflationary situation we always have $\alpha \ll 1$.

\par

As a rule of thumb, the inflaton field can be said to behave
quantum-mechanically if the correction $\mean{\df_2}$ to the mean
value is of the same order as this classical field $\varphi_\cl$. For
a more detailed argument, let us consider the calculation carried out
in Ref.~\cite{Gratton:2005bi} (see also the discussion in
Ref.~\cite{Martin:2005ir}). There, the authors compute the number of
e-folds \(\mean{N}=\int {\rm d}t\mean{H}\), which we adapt to the case
of DBI inflation as
\begin{equation}
\mean{N}=-\frac{\kappa}{2}\int_{\varphi_\ini}^{\varphi_\cl} 
{\rm d}\psi\, \frac{\mean{H}}{H_\cl'}\gamma _\cl\, .
\end{equation}
In the present case, \ie for the potential and warp factor of
Eqs.~(\ref{eq:chaotic}), this gives
\begin{equation}
\mean{N}=-\frac{\kappa}{2}\sqrt{\frac{\kappa}{6}}
\int_{\varphi_\ini}^{\varphi_\cl} 
{\rm d}\psi \,\frac{m\,\varphi_\cl\,\gamma_\cl}{H_\cl'}
\left(1+\frac{\mean{\df_2}}{\varphi_\cl}\right)\, .
\end{equation}
If $\mean{\df_2}\ll \varphi_\cl$, then $\mean{N}=N_\cl$ and the
trajectory is indeed classical, confirming our rule of thumb. If we
work out the above condition ignoring unimportant numerical factors,
we find that
\begin{equation}
\frac{\mean{\df_{2}}}{\varphi_\cl}\sim \epsilon _1\,\frac{Q^2}{T_{_{\rm
CMB}}^2}\ll 1\, .
\end{equation}
We conclude that the stochastic effects do not play any important r\^ole
in the model with $\beta=0$ and, therefore, that eternal inflation is
not possible. This conclusion, although obtained with a different
method, is in agreement with Ref.~\cite{Chen:2006hs}. In view of this
result, we do not discuss the domain of validity of the perturbative
approach in the $\beta=0$ case, but instead turn straight to the case
of a CKS potential with constant term.

\subsection{Chaotic Klebanov Strassler Inflation with a Constant Term}
\label{subsec:chaoticconst}

\begin{figure*}
\includegraphics[width=0.45\textwidth,clip=true]{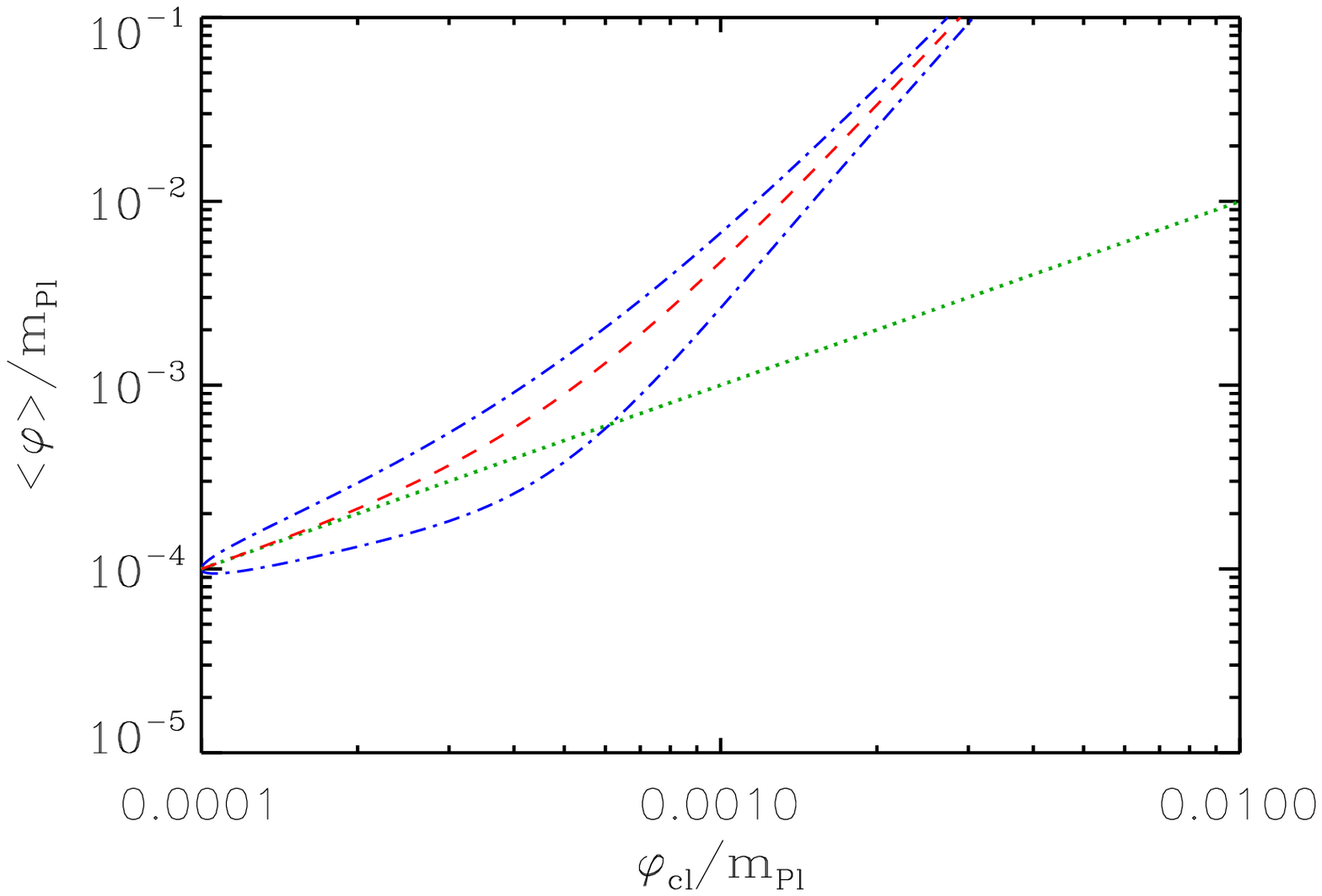}
\includegraphics[width=0.45\textwidth,clip=true]{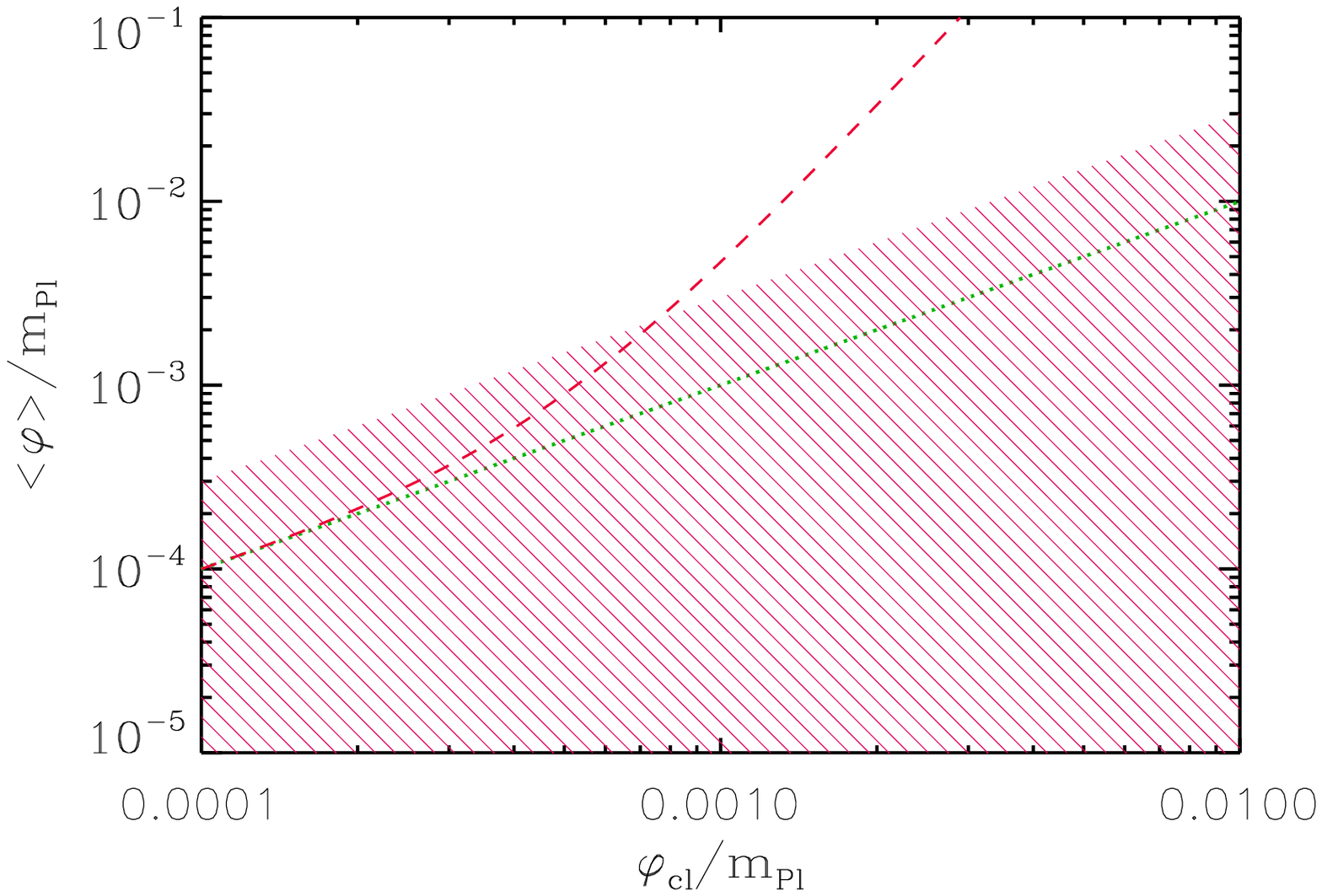}
\caption{Evolution of the (quantum) scalar field in the IR case (hence
  inflation proceeds from left to right) for $\alpha =38$, $\beta
  =3.7$ and $m\simeq 2.19\times 10^{-7}\mpl$, as implied by the COBE
  normalization. Note that the choice of the parameters is such that
  the condition $\beta/\alpha \ll 1$ is valid. For this example, the
  spectral index and the running of the model are respectively given
  by $\nS-1\simeq -0.11$ and $\alpha _{_{\rm S}}\simeq 0.0023$, as it
  can be shown using the results of Ref.~\cite{Lorenz:2008et}. The
  initial condition is $\varphi_\ini=10^{-4}\mpl$. The green dotted
  line represents the classical evolution without the quantum
  effects. The red dashed line represents the mean value of the
  quantum scalar field, namely $\varphi_\cl +\mean{\df_{2}}$ while the
  two blue dashed dotted lines on both side of the mean are
  $\varphi_\cl+\mean{\df_{2}}\pm \sqrt{\mean{\df_{1}^2}}$. On the
  right panel, the hatched region represents the region where the
  perturbative treatment used in this article is valid. For the
  parameters chosen here, the perturbative approach breaks down at
  $\varphi_\cl\simeq 0.001\mpl$.}
\label{fig:ir}
\end{figure*}

\begin{figure*}
\includegraphics[width=0.45\textwidth,clip=true]{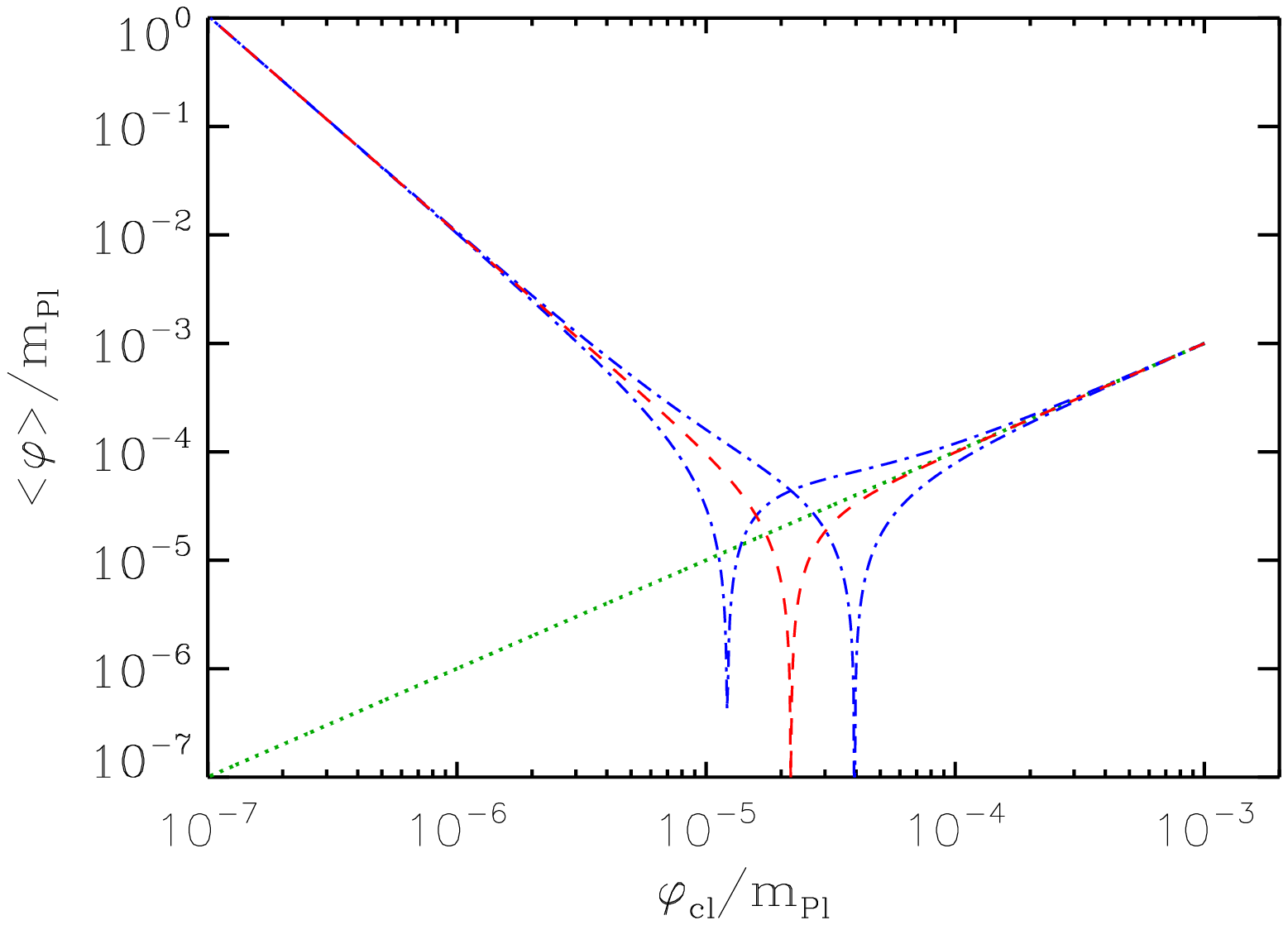}
\includegraphics[width=0.45\textwidth,clip=true]{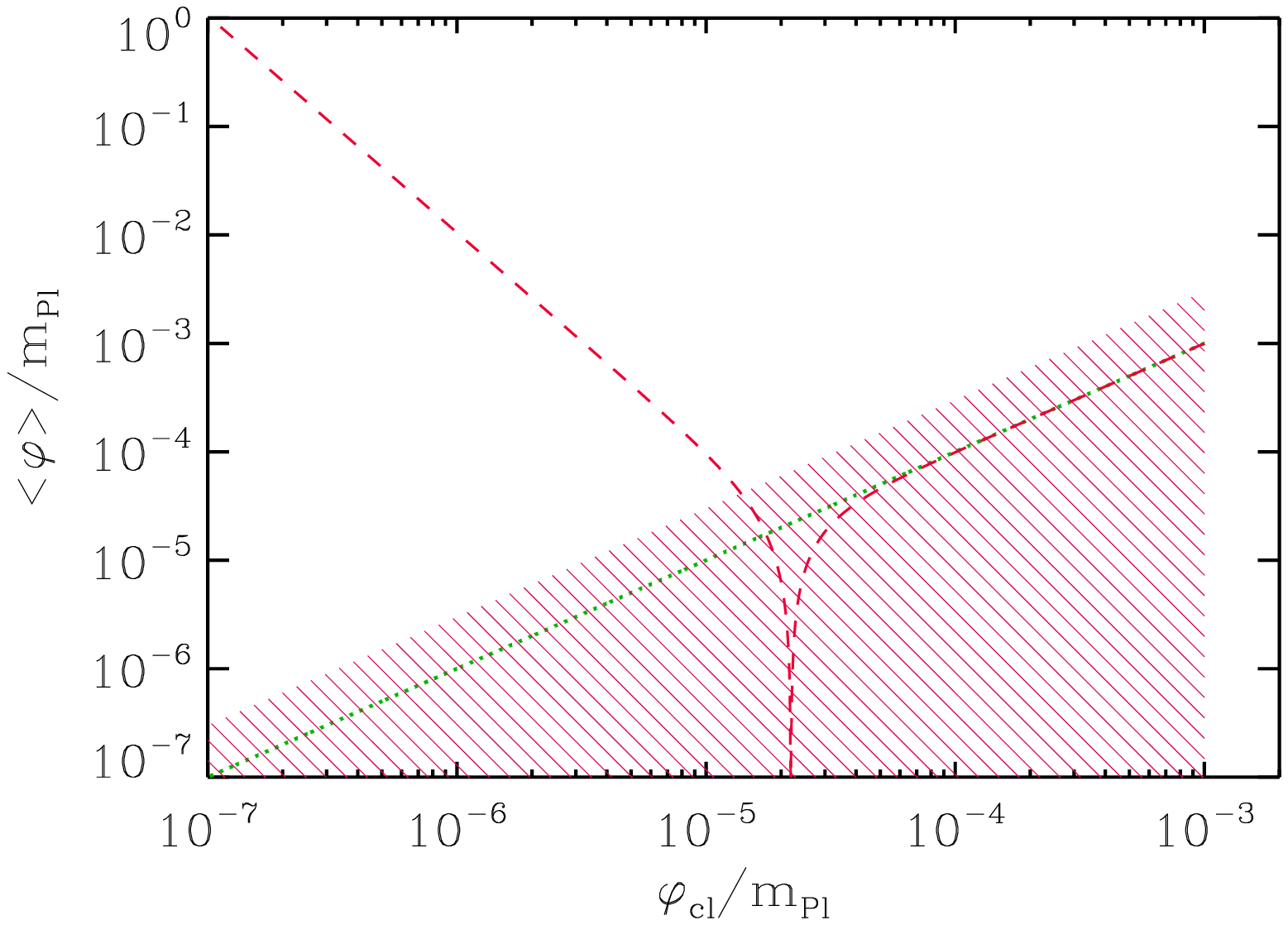}
\caption{Evolution of the (quantum) scalar field in the UV case (hence
  inflation proceeds from right to left) for $\alpha =38$, $\beta
  =3.7$ and $m\simeq 2.19\times 10^{-7}\mpl$ as implied by the COBE
  normalization (this case corresponds to Fig.~9 of
  Ref.~\cite{Lorenz:2008et}). For this example, the spectral index and
  the running of the model are respectively given by $\nS-1\simeq
  0.11$ and $\alpha _{_{\rm S}}\simeq -0.0023$. The initial condition
  is $\varphi_\ini=10^{-3}\mpl$. The conventions are analogous to
  those in Fig.~\ref{fig:ir}. The fact that $\mean{\df_{2}}_{_{\rm
      UV}}<0$, see Eq.~(\ref{eq:meanapproxuv}), and, therefore
  $\mean{\varphi}<\varphi _\cl$, can be clearly observed in the above
  plot (red dashed line). On the right panel, the hatched region
  indicates the domain of validity of the perturbative approach.}
\label{fig:uv}
\end{figure*}

As a second example, we keep $T(\phi)$ as in Eq.~(\ref{eq:chaotic}),
but take $\beta\neq0$, so that the potential is given by
\begin{equation}
\label{eq:pot}
V(\phi)=V_{0}+ \dfrac{\varepsilon}{2}\,m^{2}\phi^{2}\, .
\end{equation}
With this potential, the Lorentz factor [compare
Eq.~(\ref{eq:gammaCKS})] is given by
\begin{equation}\label{eq:gammawconst}
\gamma=\left(\frac{\mpl}{\phi}\right)^2
\sqrt{\left(\frac{\phi}{\mpl}\right)^4+\frac{1}{\alpha}
\frac{\left(\phi/\mpl\right)^2}
{2\beta +\varepsilon\left(\phi/\mpl\right)^2}}
\end{equation}
Then, in order to determine the corrections to the variance
$\mean{\df_{1}^{2}}$ and mean value $\mean{\df_{2}}$, one has to
compute the kernel of the integral in Eq.~(\ref{eq:dbidf1}). One
obtains
\begin{eqnarray}
\frac{H\gamma}{H'} &=& \frac{2\varepsilon}{m^2\phi^2}
\Biggl\{\left(V_0+\frac{\varepsilon}{2}m^2\phi^2\right)
\Biggl[\frac{V_0\mpl^2}{2\alpha \beta} \nonumber \\ & & 
+\phi^2\left(V_0+\frac{\varepsilon}{2}m^2\phi^2\right)\Biggr]\Biggr\}^{1/2}\, .
\end{eqnarray}
Unfortunately, the (third power of this) expression is too complicated
to perform the integral Eq.~(\ref{eq:dbidf1}) exactly. However, while
the constant term of the potential Eq.~(\ref{eq:pot}) dominates
($\beta\gg1$), it is legitimate to approximate $V_0+\varepsilon
m^2\phi^2/2$ by $V_0$. In this case, the calculation of
Eq.~(\ref{eq:dbidf1}) can be done and leads to
\begin{eqnarray}
\label{eq:varcksconst}
\mean{\df_{1}^{2}}&=& -\frac{32\varepsilon}{15}\frac{m^4\varphi_\cl^2}
{V_0\gamma_\cl ^2}\alpha \beta ^4\Biggl\{\Biggl[\frac{1+2\alpha \beta 
\left(\varphi_\ini/\mpl\right)^2}{2\alpha \beta 
\left(\varphi_\ini/\mpl\right)^2}
\Biggr]^{5/2}\nonumber \\
& & -\Biggl[\frac{1+2\alpha \beta 
\left(\varphi _\cl/\mpl\right)^2}{2\alpha \beta 
\left(\varphi_\cl /\mpl\right)^2}
\Biggr]^{5/2}\Biggr\}\, .
\end{eqnarray}
In the same limit, one can also estimate $\mean{\df_{2}}$. The result
reads
\begin{eqnarray}
\label{eq:meancksconst}
& & \mean{\df_{2}}= \frac{4\varepsilon\beta ^2}{3}
\left(\frac{m}{\mpl}\right)^2
\frac{\varphi_\cl}{\gamma_\cl}
\Biggl[\left(\frac{\mpl}{\phi_\ini}\right)^2\gamma _\ini^2
\nonumber \\ & &
-\left(\frac{\mpl}{\varphi_\cl }\right)^2\gamma _\cl^2\Biggr]
+\left[1+\frac{1}{2\alpha \beta}
\left(\frac{\mpl}{\varphi_\cl}\right)^2
\frac{1}{\gamma _\cl^2}
\right]\frac{\mean{\df_{1}^{2}}}{2\varphi_\cl}\, ,
\nonumber \\
\end{eqnarray}
where the explicit expression of $\mean{\df_{1}^2}$ is given in
Eq.~(\ref{eq:varcksconst}). The evolution of $\mean{\df_{1}^2}$ and
$\mean{\df_{2}}$ in the IR and UV cases is represented in
Figs.~\ref{fig:ir} and~\ref{fig:uv}.

\par

Finally, we have to estimate the volume effects. For this purpose, we
have to compute the correction to the mean value of the field given by
Eq.~(\ref{eq:voleffects}). In our approximation, one has $H^4\gamma
^3/(H')^3 \simeq \sqrt{\kappa V_0/3}\left[H^3\gamma
  ^3/(H')^3\right]$. This means that one can express the integral of
Eq.~(\ref{eq:voleffects}) in terms of the integral appearing in
Eq.~(\ref{eq:dbidf1}), \ie in terms of $\mean{\df_{1}^2}$. After
carrying out this calculation, one obtains $3I^{T}J \simeq
0$. Therefore, at first order in our approximation, volume effects
simply are absent. In fact, one could have guessed this result from
the very beginning: at first order, the Hubble parameter is given by
$H\simeq \sqrt{\kappa V_0/3}$ which is a constant. Looking at
Eq.~(\ref{eq:volume}), we see that, in this case, the weight related
to the volume can be taken outside the integrals in the numerator and
denominator. Hence, the corresponding contributions cancel out and, at
this order, there is no volume effect.

\par

For the same reason, there is no difference between the correlation
functions computed with the time variable $t$ 
or with the e-fold variable $N$,
see for instance Eqs.~(\ref{eq:dbidf1})
and~(\ref{eq:meandf1efold}). This illustrates the fact that the
difference between the two approaches can be important only if the
Hubble parameter evolves significantly. This is the case for chaotic
inflation~\cite{Finelli:2008zg,Finelli:2010sh} (although for the DBI
version of the chaotic model studied in the last subsection, the
corrections were only logarithmic) but not for the model under
scrutiny here.

\par

Before investigating in more detail the previous result, let us
establish its domain of validity. It was shown in
Ref.~\cite{Martin:2005hb} that one can trust the perturbative
stochastic treatment as long as the mean value of the inflaton field
is such that $\mean{\varphi}\in \left[\varphi_\cl -\left\vert \Delta
    \varphi_{\rm min}\left(\varphi _\cl\right)\right \vert ,
  \varphi_\cl +\Delta \varphi_{\rm max}\left(\varphi
    _\cl\right)\right]$ where $\Delta \varphi_{\rm min}$ and $\Delta
\varphi_{\rm max}$ can be found from two conditions. The first of
those reads
\begin{eqnarray}
\max_{x\in[\varphi_{\rm cl},\varphi_{\rm cl}
+\Delta\varphi(\varphi_{\rm
cl})]}\left|\frac{H^{(4)}(x)}{6}
\Delta\varphi^{3}\right|&\ll&\left|\frac{H_{\rm
cl}'''}{2}\right|\Delta\varphi^{2},
\label{eq:cond1}
\end{eqnarray}
where $H^{(4)}$ denotes the fourth order derivative. The second
condition can be expressed as
\begin{align}
\max_{x\in[\varphi_{\rm
cl},\varphi_{\rm cl}+\Delta\varphi(\varphi_{\rm
cl})]}& \left|\frac{\left[H^{3/2}(x)\right]''}{2}\right|
\Delta\varphi^{2}\nonumber \\ & \ll \left|\left(H_{\rm
cl}^{3/2}\right)'\Delta\varphi\right|.
\label{eq:cond2}
\end{align}
These two conditions must be simultaneously satisfied and, therefore,
the tightest bounds on \(\Delta\varphi_{\rm min}\) and
\(\Delta\varphi_{\rm max}\) that follow from Eq.~(\ref{eq:cond1}) and
Eq.~(\ref{eq:cond2}), respectively, give the reliability of the
perturbative treatment. Replacing again $V_0+\varepsilon m^2\phi^2/2$
by $V_0$ where applicable, straightforward manipulations show that
Eq.~(\ref{eq:cond1}) leads to $\Delta \varphi_{\rm max}=\Delta
\varphi_{\rm min}=3\varphi_\cl$ while Eq.~(\ref{eq:cond2}) gives
$\Delta \varphi_{\rm max}=\Delta \varphi_{\rm
  min}=2\varphi_\cl$. Therefore, one concludes that the perturbative
approach is correct as long as $\mean{\varphi}\in \left[\varphi_\cl
  -2\varphi_\cl, \varphi_\cl +2\varphi_\cl\right]$. The allowed region
is represented by the hatched area in Figs.~\ref{fig:ir}
and~\ref{fig:uv}.

\par

We now return to Eqs.~(\ref{eq:varcksconst})
and~(\ref{eq:meancksconst}). These expressions can be further
simplified if we take into account the fact that the vev of the
inflaton field (measured in units of the Planck mass) must be small,
compare Eq.~(\ref{eq:volumeconstraint}). In this case, the Lorentz
factor from Eq.~(\ref{eq:gammawconst}) is given by $\gamma \simeq
\mpl/(\sqrt{2\alpha \beta}\varphi_\cl)\gg 1$ and
\begin{eqnarray}
\mean{\df_{1}^{2}}_{_{\rm IR}} &\simeq & \frac{16}{15 \sqrt{2}}
\left(\frac{m}{\mpl}\right)^2
\frac{\beta ^{3/2}}{\alpha ^{1/2}}\frac{\mpl^3
\varphi_\cl^4}{\varphi_\ini^5}\, , \\
\mean{\df_{1}^{2}}_{_{\rm UV}} &\simeq & \frac{16}{15 \sqrt{2}}
\left(\frac{m}{\mpl}\right)^2
\frac{\beta ^{3/2}}{\alpha ^{1/2}}\frac{\mpl^3}{\varphi _\cl}\, .
\end{eqnarray}
In the same way, one can also calculate the correction to the vev of
the inflaton field. One obtains
\begin{eqnarray}
\label{eq:meanapproxir}
\mean{\df_{2}}_{_{\rm IR}} &\simeq & \frac{16}{15 \sqrt{2}}
\left(\frac{m}{\mpl}\right)^2
\frac{\beta ^{3/2}}{\alpha ^{1/2}}\frac{\mpl^3
\varphi_\cl^3}{\varphi_\ini^5}\, , \\
\label{eq:meanapproxuv}
\mean{\df_{2}}_{_{\rm UV}} &\simeq & -\frac{4}{15 \sqrt{2}}
\left(\frac{m}{\mpl}\right)^2
\frac{\beta ^{3/2}}{\alpha ^{1/2}}\frac{\mpl^3}{\varphi_\cl ^2}\, .
\end{eqnarray}
Notice in particular that, in the UV case, the correction is
negative. This is confirmed in Fig.~\ref{fig:uv}.

\par

With the help of these approximations, let us estimate when the
stochastic effects are important. As discussed before, one expects the
quantum effects to play a r\^ole when $\mean{\df_{2}}/\varphi_\cl \simeq
1$. In order to use Eqs.~(\ref{eq:meanapproxir})
and~(\ref{eq:meanapproxuv}) for this purpose, we must again calculate
the COBE normalization; this will allow us to re-write the parameter
combination $m^2\beta ^{3/2}/(\mpl^2\alpha ^{1/2})$ appearing in
Eqs.~(\ref{eq:meanapproxir}) and~(\ref{eq:meanapproxuv}). In
Ref.~\cite{Lorenz:2008et}, it was shown that [see Eq.~(153)]
\begin{equation}
\left(\frac{\mpl}{\varphi _*}\right)^4
=\frac{45}{16\pi}\frac{Q^2}{T_{_{\rm CMB}}^2}\left(\frac{\mpl}{m}\right)^2
\frac{\alpha}{\beta ^2}\, ,
\end{equation}
where $\varphi_*$ is the value of the inflaton field when scales of
astrophysical interest crossed out of the DBI sound horizon. Moreover,
it was also demonstrated that $\nS-1\sim 4 \delta _1$ [see
Eq.~(155)]. Therefore, we finally arrive at
\begin{equation}
\left(\frac{m}{\mpl}\right)^2\frac{\beta ^{3/2}}{\alpha ^{1/2}}\simeq 
\frac{45\pi ^3}{4}\frac{Q^2}{T_{_{\rm CMB}}^2}\left(\nS-1\right)^4
\left(\frac{\beta}{\alpha}\right)^{3/2}\, .
\end{equation}
As a consequence, for the IR case described by
Eq.~(\ref{eq:meanapproxir}), the condition $\mean{\df_{2}}_{_{\rm
    IR}}/\varphi_\cl \gta 1$ is equivalent to
\begin{eqnarray}
\frac{\varphi_\cl}{\mpl}&\gta &\left(\frac{\sqrt{2}}{12\pi^3}\right)^{1/2}
\left(\frac{Q}{T_{_{\rm CMB}}}\right)^{-1}
\left(\nS-1\right)^{-2}
\left(\frac{\beta}{\alpha}\right)^{-3/4}\nonumber \\ & &
\times \left(\frac{\varphi_\ini}{\mpl}\right)^{5/2}\, .
\end{eqnarray}
Let us discuss this expression in detail. It is interesting to note
the dependence on the initial value of the field $\varphi_\ini$. The
smaller $\varphi_\ini$ is, the sooner the influence of quantum effects
sets in. (Recall that $\varphi_\ini$ is bounded from below by the
bottom of the throat, $\varphi_\ini>\phi_{0}$, and that the inflaton
field value in this scenario increases as the $D3$-brane climbs out of
the throat.) For $\varphi_\ini=10^{-3.5}\mpl $, corresponding to the
left panel in Fig.~\ref{fig:ir}, one obtains $\varphi_\cl/\mpl \gta
0.008$ in good agreement with the plot. For an exemplary initial field
value of $\varphi_\ini=10^{-4}\mpl $, one has $\varphi_\cl/\mpl \gta
0.0005$. Therefore, if the brane starts its evolution deep inside the
throat, then the stochastic effects are dominant.

\par

For the UV case of Eq.~(\ref{eq:meanapproxuv}), one can repeat the
same discussion.  One finds that this case the limit is given by
\begin{eqnarray}
\frac{\varphi_\cl}{\mpl}&\lta &\left(\frac{3\pi^3}{\sqrt{2}}\right)^{1/3}
\left(\frac{Q}{T_{_{\rm CMB}}}\right)^{2/3}
\left(\nS-1\right)^{4/3}
\left(\frac{\beta}{\alpha}\right)^{1/2}\nonumber \, . \\
\end{eqnarray}
Contrary to the IR case, we see that there is no dependence on the
initial conditions anymore. For the parameters in Fig.~\ref{fig:uv},
one obtains $\varphi_\cl \lta 2.19\times 10^{-5}\mpl$ in good
agreement with the plot. In addition, one can also check, by comparing
the two panels of Fig.~\ref{fig:uv}, that this value does not change
much when $\varphi _\ini$ is modified. The conclusion is that in the
UV case, the stochastic effects only play an important r\^ole when the
brane is approaching the bottom of the throat, \ie towards the end of
brane inflation in its UV incarnation.

\section{Stochastic DBI Inflation and the Finite Size of Extra
  dimensions }
\label{sec:finitesize}

We have established that, in the model of Eq.~(\ref{eq:pot}), the
stochastic effects can be dominant in a relevant and realistic
inflationary regime. Our next goal is to compute how the (classical)
behavior of the inflaton field is modified in the presence of
stochastic noise. In principle, the above considerations already
answer this question since we have computed in detail the time
evolution of $\mean{\df_2}$ and $\mean{\df_1^2}$. However, it is clear
from Figs.~\ref{fig:ir} and~\ref{fig:uv} that a first issue is the
limited validity of the perturbative regime.  Outside the hatched
region in the right-hand panels of Figs.~\ref{fig:ir}
and~\ref{fig:uv}, one can no longer follow the evolution of the
quantum field. In particular, it seems obvious that the regime of
eternal inflation cannot be described in this framework, even if we
have established earlier that it is likely to exist. However, even
letting aside the issue of validity, there is another, much worse
problem that renders our previous treatment highly unsatisfactory. The
probability density functions given by Eqs.~(\ref{eq:Pc})
and~(\ref{eq:volume}) can \emph{a priori} extend into the range where
$\varphi<\phi _0$ (possibly even into the range
$\varphi<0$!). Clearly, since the inflationary scenario at hand is
built on the notion of the inflaton as the KS throat's renormalized
radial coordinate $r_{0}<r<r_{\UV}$, a field value $\varphi<\phi _0$
is inconsistent. To our knowledge, all works on stochastic inflation
in brane inflation published so far are subject to this issue.  In
fact, we face here a deep conceptional challenge: in brane inflation,
the finite size of the extra dimensions plays a fundamental r\^ole. 
(In Ref.~\cite{Lorenz:2007ze}, for example, it was shown how the finite
size condition can be exploited to cut down the allowed parameter
space at the effective field theory level.) How can this crucial
importance of the geometric restrictions be implemented into the
description of stochastic DBI inflation?

\par

We propose to address this issue by introducing a (reflecting or
absorbing) wall at $\phi _0$, following the reasoning of
Ref.~\cite{Chandrasekhar:1943ws}. (A similar technique has been used
in Ref.~\cite{Martin:2004ba} in order to study the quantum behavior of
the quintessence field. Analogously, the approach we present below 
may be applied to the case of ``DBI-essence'' 
introduced in Ref.~\cite{Martin:2008xw}.) 
This wall at the bottom of the throat then marks the 
``end of the world'' as imposed by the finite string geometry 
and prevents the stochastically corrected value of
$\varphi $ from reaching values smaller than $\phi _0$. 
Let us now explore the consequences of this proposal.

\par

Let us recall from Refs.~\cite{Chandrasekhar:1943ws,Martin:2004ba}
that, if we start with a normalized distribution $P(x)$ for the
variable $x$, \ie $\int _{-\infty}^{+\infty }P(x){\rm d}x=1$, then the
normalized probability density function in the presence of a
reflecting wall at $x=a$ is given by $P(x)+P(2a-x)$, provided that
$P(2a-x)$ is also a solution of the relevant (approximate)
Fokker Planck equation. (This is the case for a Gaussian PDF.) One can
check explicitly that this distribution is correctly normalized,
$\int _{a}^{+\infty }\left[P(x)+P(2a-x)\right]{\rm d}x=1$. As a
thought experiment, let us install a reflecting wall at the bottom of
the KS throat at the position $\phi_{0}$. Then, according to
Refs.~\cite{Chandrasekhar:1943ws,Martin:2004ba}, our new distribution
would be given by
\begin{eqnarray}
\label{eq:onewall}
P_{\rm wall}(\varphi) &=& \frac{1}{\sqrt{2\pi \mean{\df_{1}^2}}}
\Biggl\{
\exp\left[-\frac{(\varphi-\varphi_\cl-\mean{\df_2})^{2}}
{2\mean{\df_1^2}}\right]\nonumber \\ & &
+\exp\left[-\frac{(2\phi_0-\varphi-\varphi_\cl-\mean{\df_2})^{2}}
{2\mean{\df_1^2}}\right]\Biggr\}\, .
\end{eqnarray}
Notice that the wall distribution is the same with or without the volume
effects since we have shown before that $3I^{T}J \simeq 0$.

\begin{figure*}
\includegraphics[width=0.85\textwidth,clip=true]{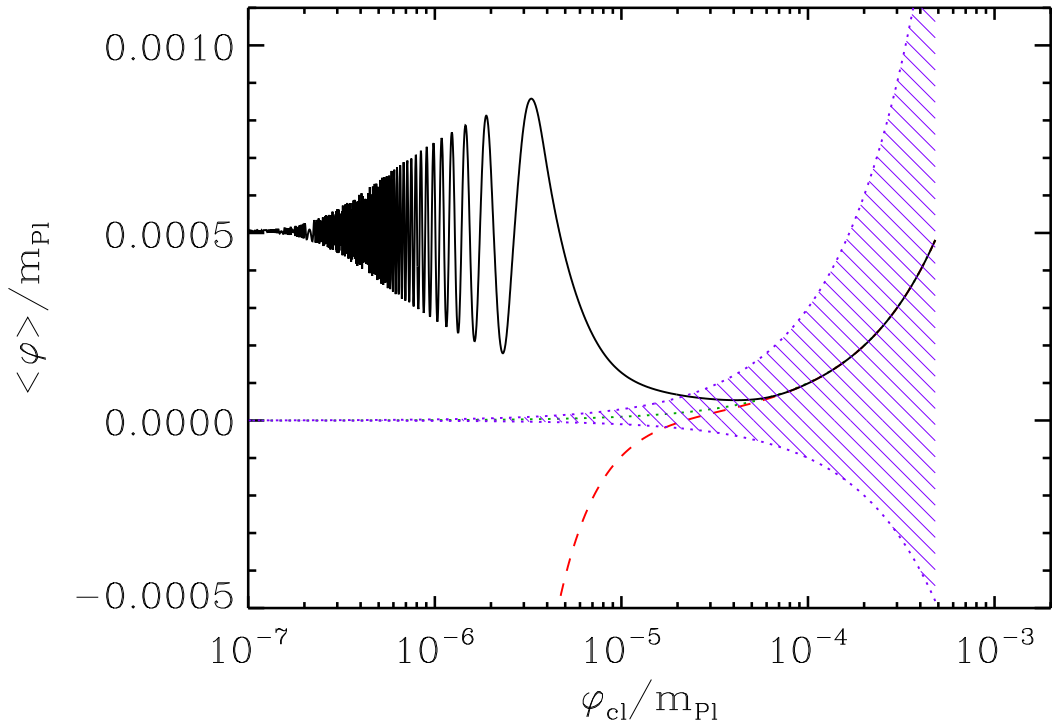}
\caption{Evolution of the (quantum) scalar field in the UV case (hence
  inflation proceeds from right to left) for $\alpha =38$, $\beta
  =3.7$ and $m\simeq 2.19\times 10^{-7}\mpl$ as implied by the COBE
  normalization (this case corresponds to Fig.~9 of
  Ref.~\cite{Lorenz:2008et}). The conventions are analogous to those
  in Fig.~\ref{fig:ir}. The black solid line represents
  $\mean{\varphi}$ given by Eq.~(\ref{eq:meanwall}) in the case where
  the two walls are located at $\phi_0=10^{-5}\mpl$ and $\phi_{_{\rm
      UV}}=10^{-3}$. The initial condition is $\phi_\ini=5\times
  10^{-4}\mpl$. This figure should be compared to
  Figs.~\ref{fig:uv}. The influence of the wall and the quantum
  trajectory is clearly visible.}
\label{fig:uvwall}
\end{figure*}

However, in the present context, when the brane reaches the bottom of
the throat, it annihilates with an anti-brane fixed at
$\phi_{0}$. Note that, while the small Coulombic contribution of the
inter-brane potential in Eq.~(\ref{eq:pot-const}) was ignored in our
calculations above, conceptually the presence of the anti-$D3$ at the
bottom of the throat is important for our present argument: therefore,
a correct physical description of this situation involves an absorbing
wall rather than a reflecting one. A more difficult question concerns 
the boundary condition to be chosen at $\phi_{_{\rm UV}}$
(where the KS throat joins the 6d bulk with unknown metric). 

\par

A more appropriate description of the situation could possibly be
given by a ``multi-throat'' scenario: for instance, one can image
that the mobile $D3$ brane travels through a 6d compact space made of
a (negligibly) small unknown bulk and two well-defined KS throats of
different depth. To account for the anti-$D3$s present at the bottom
of each throat, absorbing walls are installed at the respective
positions (or their field value counterparts, respectively).  Clearly,
the detailed treatment of such a situation will involve a high degree
of technical sophistication because a continuous description of 
warp factor and inflaton potential is needed across the 
entire 6d manifold.

\par

In order to render the problem tractable with our present means, here
we content ourselves with the assumption that a second absorbing wall
is located at $\phi_{_{\rm UV}}$. This is a good approximation for a
two-throat model in which the second throat is much ``shallower'' than
the primary inflationary throat (which is responsible for most of the
exponential expansion).  Our hope is that this toy model with
absorbing walls at both $\phi_{0}$ and $\phi_{_{\rm UV}}$ will give us
an idea about the behavior to be expected in a more realistic
case. Of course, placing a reflecting wall at $\phi_{_{\rm UV}}$ is
also technically possible but seems physically less justified at
present. Therefore, in the following, we concentrate on the case of
two absorbing walls.

\subsection{Geometrically Amended PDF}
\label{subsec:pdf}

Our goal is now to derive the PDF in presence of two absorbing
barriers, \ie the equivalent of Eq.~(\ref{eq:onewall}) but now with 
the two boundary conditions
\begin{equation}
\label{eq:bc}
P(\varphi=\phi_0)=P(\varphi=\phi_{_{\rm UV}})=0.
\end{equation}
There are two ways to derive the corresponding PDF. The first one is
based on the method of images, see \eg 
Refs.~\cite{Sommerfeld:ED, Sommerfeld:PDEs}. We
start with the unrestricted PDF $P_{\rm g}(\varphi-\phi_{\rm mean})$,
defined in Eq.~(\ref{eq:Pc}), centered at $\phi_{\rm mean}\equiv
\varphi_\cl+\mean{\df_2}$. In order to ``kill'' the contribution of
$P_{\rm g}$ at $\phi_{_{\rm UV}}$, we introduce another ``source''
located at $\phi_{_{\rm UV}}^{(1)}\equiv 2\phi_{_{\rm UV}}-\phi_{\rm
  mean}$ (this source is the ``image'' of $\phi_{\rm mean}$ with
respect to the wall located at $\phi_{_{\rm UV}}$) and weighed with a
minus sign. In the same manner, in order to remove the contribution of
$P_{\rm g}$ at $\phi_0$, we introduce a second image located at
$\phi_0^{(1)}\equiv 2\phi_0-\phi_{\rm mean}$ still weighed with a
minus sign (this source is the image of $\phi_{\rm mean}$ with respect
to the wall located at $\phi_0$). Therefore, the new PDF is given by
\begin{eqnarray}
& & P_{\rm g}(\varphi-\phi_{\rm mean})-
P_{\rm g}\left(\varphi-\phi_{_{\rm UV}}^{(1)}\right)
-P_{\rm g}\left(\varphi-\phi_0^{(1)}\right)\nonumber \\
&=&
P_{\rm g}(\varphi-\phi_{\rm mean})-P_{\rm g}(\varphi-2\phi_{_{\rm UV}}
+\phi_{\rm mean})\nonumber \\ & & -P_{\rm g}(\varphi-2\phi_0
+\phi_{\rm mean}).\label{eq:Pg}
\end{eqnarray}
However, this new distribution function does not yet satisfy the
boundary conditions~(\ref{eq:bc}) because the source $\phi_{_{\rm
    UV}}^{(1)}$ now gives a contribution at $\phi_0$. In the same
manner, the source $\phi_0^{(1)}$ gives a contribution at $\phi_{_{\rm
    UV}}$. The cure is obviously to add extra images. Therefore, we
introduce the source $\phi_{_{\rm UV}}^{(2)}$ which is the image of
$\phi_0^{(1)}$ with respect to the wall located at $\phi_{_{\rm UV}}$,
\ie
\begin{equation}
\phi_{_{\rm UV}}^{(2)}-\phi_{_{\rm UV}}=\phi_{_{\rm UV}}-\phi_0^{(1)},
\end{equation}
or
\begin{equation}
\phi_{_{\rm UV}}^{(2)}=2\phi_{_{\rm UV}}+\phi_{\rm mean}-2\phi_0.
\end{equation}
This time the PDF must be weighted with a plus sign since the new
contribution is negative. We also introduce the image $\phi_0^{(2)}$,
the image of $\phi_{_{\rm UV}}^{(1)}$ with respect to the wall located
at $\phi_0$. As a consequence, one obtains
\begin{equation}
\phi_0^{(2)}=-2\phi_{_{\rm UV}}+\phi_{\rm mean}+2\phi_0.
\end{equation}
But, as before, the new images will give new contributions on the two
walls. Clearly, in order to obtain a distribution that satisfies the
boundary conditions~(\ref{eq:bc}), this process must be repeated
\emph{ad infinitum}, \ie one must introduce an infinite number of
images. The location of $n$th source can be expressed as
\begin{eqnarray}
\phi_0^{(2n)}&=&\phi_{\rm mean}+2n\left(\phi_0-\phi_{_{\rm UV}}\right), \\
\phi_{_{\rm UV}}^{(2n)}&=&\phi_{\rm mean}+2n\left(\phi_{_{\rm UV}}-\phi_0
\right), \\
\phi_0^{(2n+1)}&=&2\phi_0-\phi_{\rm mean}
+2n\left(\phi_0-\phi_{_{\rm UV}}\right), \\
\phi_{_{\rm UV}}^{(2n+1)}&=&2\phi_{_{\rm UV}}-\phi_{\rm mean}
+2n\left(\phi_{_{\rm UV}}-\phi_0
\right).
\end{eqnarray}
Note the distinction between the positions of the even- and 
odd-numbered sources ($n=0,1,2\dots$). 
Then, the complete distribution is obtained by taking
the sum over all $n$ contributions with their appropriate 
respective signs, \ie a minus for the
``odd sources'' and a plus for the ``even sources''. The final
expression is given by
\begin{widetext}
\begin{eqnarray}
P(\varphi)&=& P_{\rm g}\left(\varphi -\phi_{\rm mean}\right)
-\sum_{n=0}^{\infty}
\left[P_{\rm g}\left(\varphi -\phi_0^{(2n+1)}\right)
+P_{\rm g}\left(\varphi -\phi_{_{\rm UV}}^{(2n+1)}\right)\right]
+\sum_{n=1}^{\infty}
\left[P_{\rm g}\left(\varphi -\phi_0^{(2n)}\right)
+P_{\rm g}\left(\varphi -\phi_{_{\rm UV}}^{(2n)}\right)\right]\nonumber\\
&=& 
\frac{1}{\sqrt{2\pi \mean{\df_{1}^2}}}
\sum _{n=-\infty}^{\infty}\left(
\exp\left\{-\frac{\left[\varphi-\phi_{\rm mean}
+2n\left(\phi_{_{\rm UV}}-\phi_0\right)\right]^{2}}
{2\mean{\df_1^2}}\right\}
-\exp\left\{-\frac{\left[\varphi+\phi_{\rm mean}-2\phi_0
+2n\left(\phi_{_{\rm UV}}-\phi_0\right)\right]^{2}}
{2\mean{\df_1^2}}\right\}\right). \nonumber \\
\label{eq:Ptwowalls}
\end{eqnarray}
This formula is very general: it gives the PDF for any Gaussian
process (characterized by the variance $\mean{\df_1^2}$ and mean value
$\mean{\df_2}$) that is restricted between two absorbing barriers.

\par

As mentioned above, there exists another way to reach the above
result. This second method consists in starting from the first line of
Eq.~(\ref{eq:Pc}), \ie the definition
$P=\mean{\delta(\varphi-\varphi[\xi])}$, but with a Dirac
$\delta$-function compatible with the boundary conditions of
Eq.~(\ref{eq:bc}). The corresponding representation of the Dirac
$\delta$-function is given by
\begin{eqnarray}
\delta (\varphi-\varphi')&=&\frac{2}{\phi_{_{\rm UV}}-\phi_0}
\sum _{n=1}^{\infty}
\sin\left(n\pi\frac{\varphi-\phi_0}{\phi_{_{\rm UV}}-\phi_0}\right)
\sin\left(n\pi\frac{\varphi'-\phi_0}{\phi_{_{\rm UV}}-\phi_0}\right).
\end{eqnarray}
This is a $\delta$-function in the sense that
$\int_{\phi_{0}}^{\phi_{_{\rm
      UV}}}\dd\varphi\,\delta(\varphi-\varphi')\,f(\varphi)=f(\varphi')$
for any function $f(\varphi)$ and any
$\varphi'\in[\phi_{0},\phi_{_{\rm UV}}]$.  From this expression, it is
straightforward to establish that
\begin{eqnarray}
\label{eq:Pdeltainter}
\mean{\delta(\varphi-\varphi[\xi])}&=&
\frac{1}{\phi_{_{\rm UV}}-\phi_0}
\sum _{n=1}^{\infty}\exp\biggl[-\frac12 
\left(\frac{n\pi}{\phi_{_{\rm UV}}-\phi_0}\right)^2\mean{\df_1^2}
\biggr]
\cos\biggl[\frac{n\pi}{\phi_{_{\rm UV}}-\phi_0}
(\varphi-\varphi_\cl-\mean{\df_2})\biggr]
\nonumber \\
&-&\frac{1}{\phi_{_{\rm UV}}-\phi_0} 
\sum _{n=1}^{\infty}\exp\biggl[-\frac12 
\left(\frac{n\pi}{\phi_{_{\rm UV}}-\phi_0}\right)^2\mean{\df_1^2}\biggr]
\cos\biggl[\frac{n\pi}{\phi_{_{\rm UV}}-\phi_0}
(\varphi+\varphi_\cl+\mean{\df_2}-2\phi_0)\biggr].
\end{eqnarray}
\end{widetext}
Then, using the identity
\begin{equation}
\sum _{n=-\infty}^{\infty}{\rm e}^{-(\varphi+n\ell)^2}
=\frac{\sqrt{\pi}}{\ell}+\sum _{n=1}^{\infty}
\frac{2\sqrt{\pi}}{\ell}{\rm e}^{-\left(\frac{n\pi}{\ell}\right)^2}
\cos\left(\frac{2n\pi \varphi}{\ell}\right),
\end{equation}
it is easy to prove that Eq.~(\ref{eq:Pdeltainter}) is in fact exactly
Eq.~(\ref{eq:Ptwowalls}). This PDF is the main result of this paper:
it is applicable to any model where the range of variation of the
stochastic inflaton field is limited by two absorbing walls. Finally,
let us notice that the distribution~(\ref{eq:Ptwowalls}) is not
normalized as is expected since the two boundary conditions are
absorbing walls.

\subsection{Inflaton Between Two Absorbing Walls}
\label{subsec:behaviourthroat}

In order to study the behavior of the inflaton field in a KS throat
with its boundaries marked by absorbing walls, we now calculate the
mean field value $\mean{\varphi}$ using the geometrically consistent
PDF (\ref{eq:Ptwowalls}) obtained in the previous Section. It is given
by the following expression
\begin{equation}
\label{eq:meanwall}
\mean{\varphi}=\frac{1}{\mathscr{N}}
\int _{\phi_0}^{\phi_{_{\rm UV}}}{\rm d}\varphi\,
P(\varphi)\,\varphi\,,
\end{equation}
where the ``normalization'' $\mathscr{N}$ can 
be calculated from
\begin{equation}
{\mathscr{N}}=\int _{\phi_0}^{\phi_{_{\rm UV}}}
P(\varphi){\rm d}\varphi.
\end{equation}
Lengthy but straightforward calculations show that 
\begin{eqnarray}
\label{eq:intwall}
& &\int _{\phi_0}^{\phi_{_{\rm UV}}}
P(\varphi)\varphi{\rm d}\varphi =\sum _{n=1}^{\infty}
\frac{2}{n\pi}\exp\left[-\frac{n^2\pi^2\mean{\df_1^2}}
{2\left(\phi_{_{\rm UV}}-\phi_0\right)^2}\right]\nonumber \\ & & \times
\sin\left[\frac{n\pi\left(\phi_0-\phi_{\rm mean}\right)}
{\phi_{_{\rm UV}}-\phi_0}\right]\left[\phi_{_{\rm UV}}\cos(n\pi)-\phi_0\right]\,,
\end{eqnarray}
while the normalization $\mathscr{N}$ is given by
\begin{eqnarray}
  \label{eq:normwall}
  {\mathscr N}&=&\sum _{n=1}^{\infty}
  \frac{4}{n\pi}\exp\left[-\frac{n^2\pi^2\mean{\df_1^2}}
    {2\left(\phi_{_{\rm UV}}-\phi_0\right)^2}\right]\nonumber \\
  & & \times \sin\left[\frac{n\pi\left(\phi_{\rm mean}-\phi_0\right)}
    {\phi_{_{\rm UV}}-\phi_0}\right]
  \sin ^2\left(\frac{n\pi}{2}\right).
\end{eqnarray}
Finally, one can push this reasoning further to calculate the the variance 
of the field from $\mean{\varphi^{2}}$ defined as
\begin{equation}
\label{eq:meansqrdwall}
\mean{\varphi^{2}}=\frac{1}{\mathscr{N}}
\int _{\phi_0}^{\phi_{_{\rm UV}}}{\rm d}\varphi\,
P(\varphi)\,\varphi^{2}\,.
\end{equation}
Using the same techniques, one obtains
\begin{widetext}
\begin{eqnarray}
\mean{\varphi^2} &=&\frac{1}{\mathscr{N}}\sum _{n=1}^{\infty}
\frac{2}{n\pi}\exp\left[-\frac{n^2\pi^2\mean{\df_1^2}}
{2\left(\phi_{_{\rm UV}}-\phi_0\right)^2}\right] \\
&&\times \sin\left[\frac{n\pi\left(\phi_{\rm mean}-\phi_0\right)}
{\phi_{_{\rm UV}}-\phi_0}\right]
\left\{\left[\phi_0^2
-2\lmk\frac{\phi_{_{\rm UV}}-\phi_0}{n\pi}\rmk^2\right]-\cos (n\pi)\left[\phi_{_{\rm UV}}^2
-2\lmk\frac{\phi_{_{\rm UV}}-\phi_0}{n\pi}\rmk^2\right]
\right\}. \nonumber 
\end{eqnarray}
\end{widetext}

\par

It is interesting to check the consistency of the above expressions at
the initial time (when $\varphi=\varphi_\ini$).  Initially,
$\mean{\df_2}=\mean{\df_1^2}=0$ and the series in
Eqs.~(\ref{eq:intwall}) and~(\ref{eq:normwall}) can be calculated
explicitly using formulas~~(1.441.1) and~(1.441.3) of
Ref.~\cite{Gradshteyn:1965aa}. The result reads $\int
_{\phi_0}^{\phi_{_{\rm UV}}}{\rm
  d}\varphi\,P(\varphi)\,\varphi=\varphi_\ini$ and
${\mathscr{N}}=1$. As a consequence, one checks that
$\mean{\varphi}_\ini=\varphi_\ini$ as expected.

\par

The quantity $\mean{\varphi}$ is represented in Fig.~\ref{fig:uvwall}
in the case where the two walls are respectively located at
$\phi_0=10^{-5}\mpl$ and $\phi_{_{\rm UV}}=10^{-3}\mpl$. The influence
of the walls is clearly visible from a comparison of the two
curves. While the unbounded distribution artificially predicts that
the brane can be outside the throat, the ``true'' quantum trajectory
accounting for the geometric restrictions shows an interesting
behavior: as the brane is approaching the wall, its position starts
oscillating and then becomes stabilized at a value $>\phi_0$. Hence,
the presence of the wall prevents $\varphi$ from violating its
geometric limit $\phi_{0}$, \ie from going beyond the bottom of the
throat.

\par

Several words of caution are in order here. Firstly, the above
conclusion is valid for particular initial conditions. Clearly, if the
parameters are such that the stochastic correction $\mean{\df_2}$
remains small, the presence of the wall is not felt. Secondly, a much
more serious problem is that the perturbative approach is not valid in
the regime where the brane position is oscillating, \ie on the left
part of Fig.~\ref{fig:uvwall}. Recall, however, that here we are
considering the UV scenario in which the inflaton field value
decreases during inflation. Therefore, Fig.~\ref{fig:uvwall} should be
read ``from right to left'': in the classical and the perturbative
stochastic approach \emph{without the walls} described by
Eq.~(\ref{eq:Pc}) (green dotted and red dashed lines, respectively),
the brane travels towards the bottom of the throat, well within the
regime of validity of our perturbative treatment. Initially, also the
trajectory found \emph{in the presence of the walls} from
Eq.~(\ref{eq:Ptwowalls}) (black solid line) overlaps with these, but
in the vicinity of $\phi_{0}$, the ``geometry-conscious'' brane
changes direction: it starts climbing upwards in the throat again, as
can be seen in Fig.~\ref{fig:uvwall}. This turnaround occurs within
the (hatched) region where the perturbative approach is
reliable. 

\par

In addition, it seems reasonable to conjecture that the oscillatory
behavior mentioned above is real despite the fact that it occurs
outside the regime of validity of our approximation. The reason is as
follows. In fact, the regime of validity indicates where the
calculation of $\mean{\df_1^2}$ and $\mean{\df_2}$ is no longer
reliable. But it does not limit in any way the validity of
Eqs.~(\ref{eq:Ptwowalls}), (\ref{eq:intwall})
and~(\ref{eq:normwall}). Clearly, the oscillatory behavior comes from
the peculiar structure of the PDF in Eq.~(\ref{eq:Ptwowalls}) which
gives rise to the appearance of trigonometric functions in
Eqs.~(\ref{eq:intwall}) and~(\ref{eq:normwall}). As long as the
stochastic effects grow (and no matter how quantitatively the do, \ie
no matter the detailed behavior of $\mean{\df_1^2}$ and
$\mean{\df_2}$), the brane will feel the wall at some point and,
consequently, the trigonometric functions in Eqs.~(\ref{eq:intwall})
and~(\ref{eq:normwall}) will start to play a r\^ole. Hence, it is very
likely that the brane position will oscillate even if, with the
perturbative method used here, we cannot calculate the fine structure
of these oscillations.

\par

Therefore, while the results obtained above probably can not be
considered a fully realistic calculation of the quantum trajectory, we
may well take them as an indication that geometric limits are of great
importance in stochastic DBI inflation. The finite size of the extra
dimensions, modelled by the presence of the two absorbing walls,
changes the stochastic corrections to the classical field trajectory
considerably.

\par 

Finally, we have proven that, for the CKS potential with
a constant term, stochastic effects can be dominant and occur near the
bottom of the throat. The question of eternal inflation is clearly a
more complicated issue. In particular, computing the stationary
distribution in this case cannot be done using the present formalism,
even if one can argue (as we have above) that the existence of an
eternally inflating regime is likely in brane inflation.

\section{Conclusions}
\label{sec:conclusions}

We now conclude our investigation by revisiting our main results. In
this paper, we have generalized the approach of
Refs.~\cite{Martin:2005ir,Martin:2005hb} to DBI inflation models, in
which the kinetic term of the inflaton is modified through an
geometry-imposed upper limit on the field velocity. To this end, we
solved the DBI Langevin equation using a perturbative expansion in the
noise. It turns out that the results of
Refs.~\cite{Martin:2005ir,Martin:2005hb} essentially only change by
additional factors of $\gamma$, reflecting the fact that the
distinction between long and short wavelength fluctuations, which is
at the heart of the stochastic inflationary approach, now is defined
with respect to the sound (instead of the Hubble) radius scale.

\par

Our calculation yields easy expressions for the PDFs describing the
probability for the patch-averaged inflaton to assume a given value in
one Hubble domain, or in the entire Universe taking into account the
size of each averaging domain. We calculated and plotted these PDFs
for the example of chaotic Klebanov Strassler inflation, both for the
cases with and without a constant term in the potential. In the
absence of a constant term $V_{0}$, stochastic effects are found not
to alter the classical field trajectory in a significant way. However,
in both the UV ($\varepsilon=+1$) and IR ($\varepsilon=-1$)
formulation of a potential including a constant term, the quantum
behavior of the field plays a dominant r\^ole for small field values,
\ie at the bottom of the Klebanov Strassler throat. Hence, in the UV
model, where the brane moves downwards in the throat, these effects
occur at the end of inflation, whereas in the IR case, with the brane
climbing up the throat, they come into play at the very beginning.

\par

The main result of this paper is our demonstration that 
an additional subtlety arises due to the
geometric restriction which confines 
the field value of the coarse-grained field 
between $\phi_{0}<\varphi<\phi_{_{\rm UV}}$: 
to prevent the PDFs from
extending below $\phi_{0}$ (where the throat --and, consequently, 
the stringy spacetime itself-- ends), we introduced two
absorbing walls located at $\phi_{0}$ and $\phi_{_{\rm UV}}$.  This
changes the PDF's shape, and within the bounds of validity it is the
``absorbing walls probability density functions'' that must be used
to describe the stochastic behavior of the inflaton field. 
It should be interesting to put this technique to use for the  
``DBI version'' of quintessence proposed in Ref.~\cite{Martin:2008xw}.

\par

The question of existence of an eternally self-reproducing regime in
the DBI case is more complicated. We have argued that such a regime
does not exist if the potential is of the pure $m^{2}\phi^2$ type, but
that in the case of an additional constant term $V_{0}$, eternal
inflation seems possible.  Given the intrinsic validity limitations of
our present approach, a fully numerical treatment of the DBI Langevin
equation would be necessary to answer this question.

\par

Finally, we would like comment on possible brane trajectories beyond
the single throat model. Globally, the string geometric 6d background
is given by a compactified bulk Calabi Yau space (where the metric is
unknown) whose ``corners'' may comprise several (possibly generalized)
Klebanov Strassler throats with different parameters. Some of these
throats can contain additional anti-branes at their bottom, like in
the original scenario of Ref.~\cite{Kachru:2003sx}.  One could
therefore imagine that a brane starts out at the bottom of an IR
throat, moving upwards towards the bulk, then transverses a short
distance within the bulk before dropping down into a UV throat.

\par

Let us consider qualitatively the quantum effects experienced by a
brane following such a trajectory: While in the IR throat, the impact
of stochastic inflation is strongest at the very bottom, \ie quantum
effects will push the brane upwards even faster than its classical
evolution demands. Having made its way across the bulk, the brane will
descend into the UV throat first on a purely classical
trajectory. However, as it nears the bottom of the second throat, the
stochastic influence grows again, pushing the brane away from the
bottom and possibly upwards into the bulk again. At the limit, one may
therefore imagine a scenario where the brane is propelled out of a UV
throat back into the bulk every time it comes sufficiently close
to a throat's bottom.

\par

Since we ignore the formulation of the metric in the bulk, it seems
difficult to carry out the full calculation for the brane trajectory
sketched above. However, we do know how to describe the metric within
each individual throat. As a first toy model, we therefore considered
the case of a brane trapped between two absorbing walls located at one
throat's bottom and edge, respectively. This successfully prevents 
the brane from meandering ``outside of spacetime'' (below 
the bottom of the throat at $\phi_{0}$), and we have 
therefore solved the severe conceptual issue stated in 
the introduction of this paper. Near the edge of 
the throat at $\phi_{_{\rm UV}}$, the technical difficulty of 
describing the stochastic effects remains because to-date 
the 6d bulk metric remains unknown. The next step would be to
write down a joint warp factor for two throats of different depth 
which are ``glued together'' at their edges. 
Given the quantum behavior found
in this paper, our expectations might point us towards an increased
probability to find the brane near the ``throat matching
point''.

\section*{Acknowledgements}

JM would like to thank Giovanni Marozzi for useful discussions. LL is
grateful to Se\'{a}n Murray for helpful comments. This work was
partially supported by JSPS Grant-in-Aid for Scientific Research
No.~19340054 (JY), the Grant-in-Aid for Scientific Research on
Innovative Areas No.~21111006 (JY) and JSPS Core-to-Core Program
``International Research Network on Dark Energy''(JM and JY). LL is
supported by the Belgian Federal Office for Scientific, Technical and
Cultural Affairs through the Interuniversity Attraction Pole P6/11.

\begin{appendix}

\section{Slow-Roll Limit}
\label{app:srlimit}

In this short appendix, we justify the use of the equation $H^2\simeq
\kappa V/3$ in the DBI case, and present an alternative derivation of
the (classical) slow-roll DBI Klein Gordon equation. 
In the standard (non-DBI) case, a convenient tool to study the
inflationary evolution is the following hierarchy of slow-roll
parameters~\cite{Schwarz:2001vv,Leach:2002ar,Schwarz:2004tz}
\begin{equation}
\label{eq:defsrparam}
\epsilon _{n+1}=\frac{{\rm d}\ln \vert \epsilon _n \vert}{{\rm d}N}\, , 
\quad \epsilon _0\equiv \frac{H_{\rm in}}{H}\, .
\end{equation} 
Then, straightforward manipulations of the background Einstein equations 
lead to the two following equations
\begin{eqnarray}
H^{2}&=&\frac{\kappa}{3}\frac{V}{1-\epsilon_1/3}\, ,
\label{eq:srHsqrd}\\
\left(3-\epsilon_{1}+\frac{1}{2}\epsilon_{2}\right)H\dot{\phi}
&=&-V'\, ,
\label{eq:sreofm}
\end{eqnarray}
where a prime denotes a derivative with respect to the inflaton field.
In the slow-roll limit where $\epsilon_{1},\epsilon_{2}\ll1$, we
therefore obtain the (Friedmann) equation $H^2\simeq \kappa V/3$ and
the slow-roll version of the Klein Gordon equation, namely
$3H\dot{\phi}\simeq -V'$.

\par

We now derive the same equations, but for DBI inflation. In this case,
the hierarchy~(\ref{eq:defsrparam}) is no longer sufficient and, in
order to describe the evolution of the Lorentz factor $\gamma$ defined
in Eq.~(\ref{eq:gamma}), we now introduce the additional hierarchy of
DBI parameters $\delta_{i}$, the so-called ``sound flow parameters'',
see \eg Refs.~\cite{Lorenz:2008et,Lorenz:2008je}. These are defined as
\begin{equation}
\delta_{n+1}=\frac{{\rm d}\ln \vert \delta_n \vert}{{\rm d}N}\, , 
\quad \delta_0\equiv \frac{\gamma}{\gamma_\ini}\, .
\end{equation} 
To find the ``slow-roll'' version of Eqs.~(\ref{eq:Hdbi})
and~(\ref{eq:eofmdbi}) (where by ``slow-roll'' we now understand that
both of the parameter sets $\epsilon_{i},\delta_{i}\ll1$), we use the
DBI background equations and the definition of $\gamma$ in
Eq.~(\ref{eq:gamma}) to show that the following expression holds
\begin{equation}
\label{eq:replacephidotsqrd}
\dot{\phi}^{2}=\frac{2}{\kappa\gamma}\,H^{2}\,\epsilon_{1}\, , 
\end{equation}
from which we can deduce that 
\begin{equation}
\ddot{\phi}=\frac{H}{2}\dot{\phi}
\left(\epsilon_{2}-2\epsilon_{1}-\delta_{1}\right)\, .
\end{equation} 
To replace the terms proportional to $T'$ in Eq.~(\ref{eq:eofmdbi}),
we multiply the expression for $\delta_{1}$ (see
Ref.~\cite{Lorenz:2008et}) by $\dot{\phi}$ and use
Eq.~(\ref{eq:replacephidotsqrd}) to find
\begin{equation}
T'=-\frac{V'}{\gamma-1}-3H\frac{\gamma}{\gamma-1}\dot{\phi}
-H\dot{\phi}\frac{\gamma^{3}}{(\gamma^{2}-1)(\gamma-1)}\, .
\end{equation}
When we use these $\epsilon_{i},\delta_{i}$ to rewrite
Eqs.~(\ref{eq:Hdbi}) and~(\ref{eq:eofmdbi}), we therefore find
\begin{eqnarray}
& & H^{2}=\frac{\kappa}{3}\,
\frac{V}{1-2\epsilon_{1}\gamma/\left(3\gamma+3\right)}\,
 ,\\
& & \frac{\dot{\phi}}{2}\left[\epsilon_{2}-2\epsilon_{1}
+\frac{\delta_{1}}{(\gamma+1)}+\frac{3(\gamma+1)}{\gamma}\right]
\nonumber \\ & &
=-\frac{\gamma+1}{2\gamma^{2}}\frac{V'}{H}\, .
\label{eq:eofm-epsdelta}
\end{eqnarray}
As announced, in the limit where both $\epsilon_{i},\delta_{i}\ll1$,
one recovers that $H^2\simeq \kappa V/3$ and
Eq.~(\ref{eq:KG-DBIsr}). Note that in Eq.~(\ref{eq:eofm-epsdelta})
both the $\epsilon_{i}$ and $\delta_{i}$ are supposed small to obtain
Eq.~(\ref{eq:KG-DBIsr}), while it sufficient to have $\epsilon _1\ll
1$ to re-derive the slow-roll version of the Friedmann equation.

\section{Perturbed Klein Gordon Equation}
\label{app:dphiderivation}

In this appendix, we briefly sketch how to obtain the formulation
Eq.~(\ref{eq:dphifinal_dbi}) for the DBI Klein Gordon equation at the
linearly perturbed level.  For the case of a standard inflaton field,
it is well known that the Klein Gordon equation for linear
perturbations of the field $\delta\phi_{\bm k}$ and scalar
perturbations of the metric $\Phi_{\bm k}$ reads
\cite{Mukhanov:1990me}
\begin{equation}\label{eq:KGperturb}
\delta\ddot{\phi}_{\bm k}+3H\delta\dot{\phi}_{\bm k}
+\left(\frac{k^{2}}{a^{2}}+V''\right)\delta\phi_{\bm k}
=4\dot{\phi}\dot{\Phi}_{\bm k}-2V'\Phi_{\bm k}\,.
\end{equation}
One can express $V'$ and $V''$ in terms of the slow-roll 
parameters $\epsilon_{i}$ defined in the Appendix \ref{app:srlimit},
\begin{eqnarray}
V'&=&-\frac{z}{a}\,H^{2}\left(3-\epsilon_{1}
+\frac{\epsilon_{2}}{2}\right),\label{eq:Vprime}\\
V''&=&H^{2}\bigg(6\epsilon_{1}-\frac{3}{2}\,
\epsilon_{2}\nonumber\\
&&-2\epsilon_{1}^{2}-
\frac{\epsilon_{2}^{2}}{4}+\frac{5}{2}\,
\epsilon_{1}\epsilon_{2}-\frac{1}{2}\,
\epsilon_{2}\epsilon_{3}\bigg),\label{eq:V2prime}
\end{eqnarray}
where the function $z=a\dot{\phi}/H$. Moreover, we
have~\cite{Mukhanov:1990me}
\begin{equation}
\dot{\Phi}_{\bm k}=H\epsilon_{1}\frac{a}{z}\,\delta\phi_{\bm k}
-H\Phi_{\bm k}\,.\label{eq:dPhi}
\end{equation}
Therefore, using Eqs.~(\ref{eq:Vprime})-(\ref{eq:dPhi}) we can rewrite
Eq.~(\ref{eq:KGperturb}) as
\begin{eqnarray}
\label{eq:dphifinal}
& &\delta\ddot{\phi}_{\bm k}+3H\delta\dot{\phi}_{\bm k}
+H^{2}\bigg(\frac{k^{2}}
{a^{2}H^{2}}+2\epsilon_{1}
-\frac{3}{2}\,\epsilon_{2}-2\epsilon_{1}^{2}
-\frac{\epsilon_{2}^{2}}{4}\nonumber \\ & &+ 
\frac{5}{2}\,\epsilon_{1}\epsilon_{2}
-\frac{\epsilon_{2}\epsilon_{3}}{2}\bigg)\delta\phi_{\bm k}=
\frac{H^{2}z}{a}\,\Phi_{\bm k}\left(2-2\epsilon_{1}+\epsilon_{2}\right)\,.
\end{eqnarray}

\par

Note that precisely the same equation is found by working backwards
from the well-known evolution of the standard Mukhanov Sasaki variable
$v_{\bm k}$ [in conformal time $\eta$, where $\dd t=a\dd \eta$,
Eq.~(\ref{eq:vk_eofm}) simplifies to $v''_{\bm
  k}+\left(k^{2}-z''/z\right)v_{\bm k}=0$, compare Eq.~(\ref{eq:v})],
\begin{equation}\label{eq:vk_eofm}
\ddot{v}_{\bm k}+H\dot{v}_{\bm k}+
\left(\frac{k^{2}}{a^{2}}-
\frac{\ddot{z}+H\dot{z}}{z}\right)v_{\bm k}=0\,,
\end{equation}
which is defined as
\begin{equation}
v_{\bm k}=a\,\delta\phi_{\bm k}+z\Phi_{\bm k},
\end{equation}
To see this, express the time derivatives of $z$ in terms of the
$\epsilon_{i}$ parameters, and use Eq.~(\ref{eq:dPhi}) as well as its
derivative along with
\begin{eqnarray}
\frac{z}{a}\frac{k^{2}}{a^{2}}\,\Phi_{\bm k}&=&
\frac{\epsilon_{1}H^{2}}{2}\left(\epsilon_{2}-2\epsilon_{1}\right)
\delta\phi_{\bm k}-\epsilon_{1}H\,\delta\dot{\phi}_{\bm k}\nonumber\\
&&+\epsilon_{1}H^{2}\frac{z}{a}\,\Phi_{\bm k},\,,
\end{eqnarray}
which is a consequence of the Einstein equations
\cite{Mukhanov:1990me}.

\par

We now proceed to deriving the analogous equations for a scalar field
with DBI dynamics. Not surprisingly, the expressions are significantly
more involved in this case, and the perturbed DBI Klein Gordon
equation reads
\begin{widetext}
\begin{eqnarray}
& &\delta\ddot{\phi}_{\bm k}+\left\{3H\,
\frac{3-2\gamma^{2}}{\gamma^{2}}-\frac{3(\gamma^{2}-1)}{\dot{\phi}\gamma^{3}}
\left[V'+(\gamma-1)T'\right]\right\}
\delta\dot{\phi}_{\bm k}
+\biggl\{\frac{k^{2}}{a^{2}\gamma^{2}}
+\frac{V''}{\gamma^{3}}-
\frac{T''}{2\gamma^{3}}
\left(\gamma^{3}-3\gamma+2\right)
+\frac{\gamma^{2}-1}{2\gamma}\frac{T'}{T}
\nonumber \\ & &\times
\left[\frac{6H}{\gamma}\,\dot{\phi}+
\frac{3V'}{\gamma^{2}}+\frac{3T'}{\gamma^{2}}\,
\left(\gamma-1\right)\right]\biggr\}
\delta\phi_{\bm k}
=\frac{4}{\gamma^{2}}\,\dot{\Phi}_{\bm k}
\dot{\phi}+\frac{1}{\gamma^{2}}\,\Phi_{\bm k}
\bigg[-\frac{\left(\gamma^{2}-1\right)^{2}}{\gamma^{2}}
\,6H\dot{\phi}-\frac{V'}{\gamma^{3}}\,
\left(3\gamma^{4}-3\gamma^{2}+2\right)
\nonumber \\ & &
-\frac{T'}{\gamma^{3}}\left(3\gamma^{5}
-3\gamma^{4}-4\gamma^{3}+3\gamma^{2}
+3\gamma-2\right)\bigg]\,.
\label{eq:DBI-KGperturb}
\end{eqnarray}
\end{widetext}
Despite its complicated appearance, it is easy to see that this
equation reduces to Eq.~(\ref{eq:KGperturb}) upon setting
$\gamma=1$. Lengthy but straightforward calculations allow to express
each term in Eq.~(\ref{eq:DBI-KGperturb}) in terms of the
$(\epsilon_{i},\delta_{i})$ parameter hierarchy defined in Appendix
\ref{app:srlimit}, in close analogy to the expressions
(\ref{eq:Vprime}) and (\ref{eq:V2prime}) in the standard case.

\par

In addition, the DBI version of Eq.~(\ref{eq:dPhi}) reads
\begin{equation}
\dot{\Phi}_{\bm k}=H\epsilon_{1}\frac{a\gamma^{3/2}}{z}\,
\delta\phi_{\bm k}-H\Phi_{\bm k}\,,
\label{B10}
\end{equation}
now with $z=a\gamma^{3/2}\dot{\phi}/H$. Combining all of these
ingredients, one arrives at Eq.~(\ref{eq:dphifinal_dbi}) in Section
\ref{subsec:stochainf}, which is equivalent to
Eq.~(\ref{eq:dphifinal}) for the case of standard inflation in which
$\gamma=1$, $\delta_{i}=0$.

\par

Again, Eq.~(\ref{eq:dphifinal_dbi}) could also have been obtained from
\begin{equation}
\ddot{v}_{\bm k}+H\dot{v}_{\bm k}+
\left(\frac{k^{2}}{a^{2}\gamma^{2}}-
\frac{\ddot{z}+H\dot{z}}{z}\right)v_{\bm k}=0\,,
\end{equation}
where in the DBI case, too, the definition of the Mukhanov-Sasaki
variable is
\begin{equation}
v_{\bm k}=a\,\gamma^{3/2}\,\delta\phi_{\bm k}+z\Phi_{\bm k}\, ,
\end{equation}
but with the new $z$ defined below Eq.~(\ref{B10}). Here we have used
\begin{eqnarray}
\frac{z}{a\gamma^{3/2}}\frac{k^{2}}{a^{2}\gamma^{2}}\,\Phi_{\bm k}
&=&\frac{\epsilon_{1}H^{2}}{2}\left(\epsilon_{2}-2\epsilon_{1}
-\delta_{1}\right)\delta\phi_{\bm k}-\epsilon_{1}H\,
\delta\dot{\phi}_{\bm k}\nonumber\\
&&+\epsilon_{1}H^{2}\frac{z}{a\gamma^{3/2}}\Phi_{\bm k},
\end{eqnarray}
as found from the Einstein equations with DBI scalar matter.
\end{appendix}

\bibliography{DBIbibliography}

\end{document}